\documentclass[journal]{IEEEtran}
\usepackage{amsmath,bm,amssymb,xcolor,multirow,mathtools,hyperref,nameref,relsize,amsthm,placeins,boldline,babel,enumitem}

\usepackage{graphicx,titlesec}
\newtheorem{theorem}{Theorem}
\newtheorem{assump}{Assumption}
\newtheorem{lemma}{Lemma}
\usepackage{float}
\usepackage{caption}
\usepackage{tikz}
\usetikzlibrary{shapes,arrows}
\usepackage{multicol}
\usepackage{setspace}

\usepackage{subfig}
\usetikzlibrary{fit}
\usepackage[top=2cm, bottom=2cm, left=2cm, right=2cm]{geometry}

\usepackage[utf8]{inputenc}

\usepackage{authblk}
\newcolumntype{C}{>{\centering\arraybackslash}X|}
\ifCLASSINFOpdf
\else
\fi
\begin{document}
	\title{On estimating parameters of a multi-component Chirp Model with equal chirp rates}
	\author{Abhinek~Shukla, ~Debasis~Kundu,   ~Amit~Mitra, and~Rhythm~Grover
		\thanks{A. Shukla, D. Kundu  and A. Mitra are with Department of Mathematics and Statistics, Indian Institute of Technology Kanpur, Kanpur - 208016, India.}
		\thanks{R. Grover is with Mehta Family School of Data Science and Artificial Intelligence, Indian Institute of Technology Guwahati, Assam-781039, India.}
		
		\thanks{(Corresponding author: Abhinek Shukla, email: abhushukla@gmail.com)}}
	
	\maketitle
	
	\begin{abstract}
		Multi-component chirp signal models with equal chirp rates appear in various radar applications, e.g., synthetic aperture radar, echo signal of a rapid mobile target, etc. Many sub-optimal estimators have been developed for such models,  however, these suffer from the problem of either identifiability or error propagation effect. In this paper, we have  developed theoretical properties of the least squares estimators (LSEs) of the parameters of multi-component chirp model with equal chirp rates, where the model is contaminated with linear stationary errors. We also propose two computationally efficient estimators as alternative to LSEs, namely sequential combined estimators and sequential plugin estimators. Strong consistency and asymptotic normality of these estimators have been derived. Interestingly, it is  observed that  sequential combined estimator of the  chirp rate parameter is asymptotically efficient. Extensive numerical simulations have been performed, which validate satisfactory computational and  theoretical performance of all three estimators.  {We have also analysed a simulated radar data with the help of our proposed estimators of multi-component chirp model with equal chirp rates, which performs efficiently in recovery of inverse synthetic aperture radar (ISAR) image of a target from a noisy data.}
	\end{abstract}
	\text{ \\~\\~\\ }{\bf Keywords:}
	Consistency, asymptotic normality, equal chirp rates, multi-component chirps, stationary linear process, least squares estimators, sequential procedures.
	
	\section{Introduction}
\begin{figure}
		\centering
		\subfloat[{\textit{Five-component chirp model with equal chirp rates.} }
		]{\label{fig:chirpfive}{\includegraphics[width=0.92\linewidth]{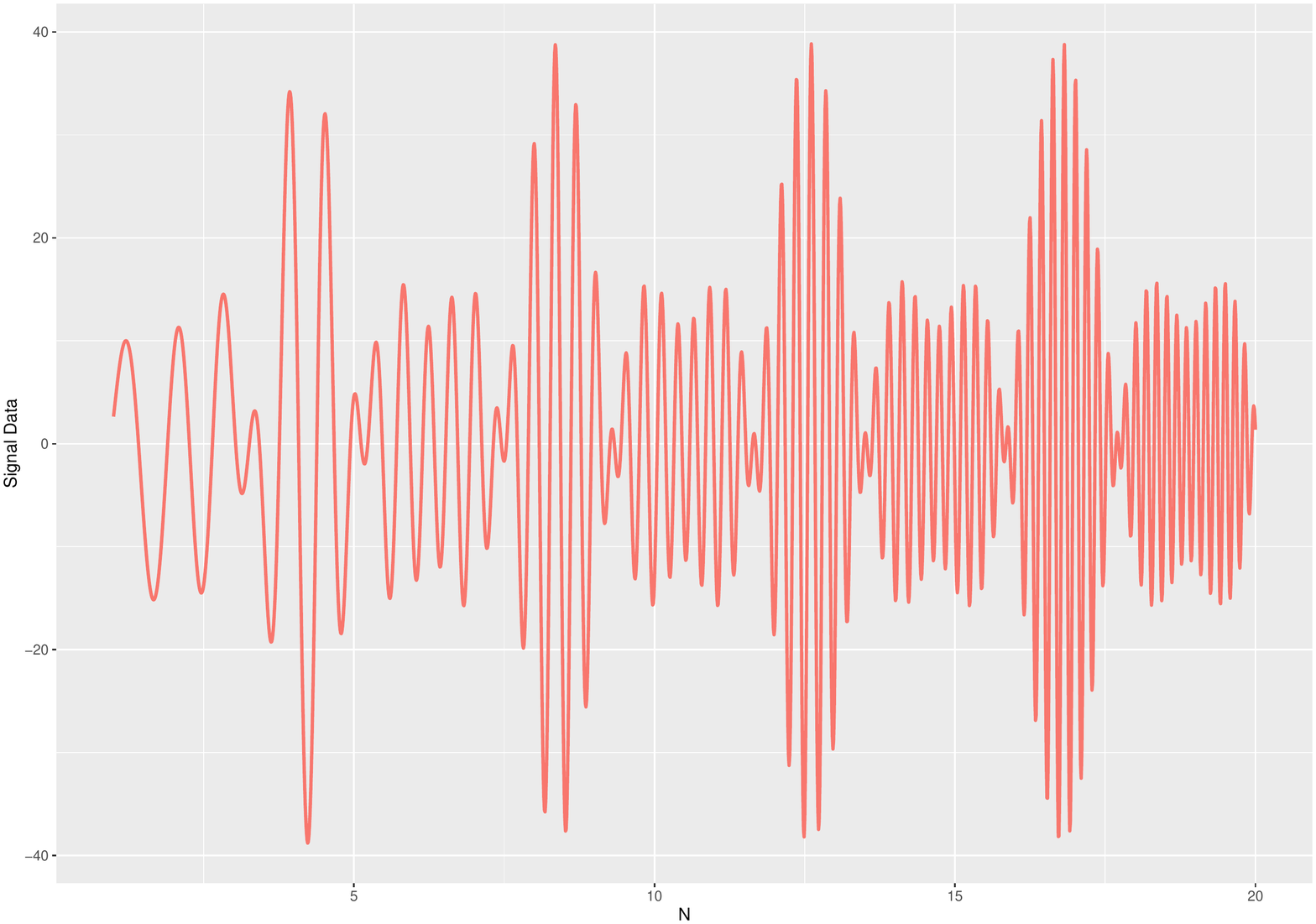}}}\\
		\subfloat[{\textit{Ten-component chirp model with equal chirp rates.}}]{\label{fig:chirpten}{\includegraphics[width=0.92\linewidth]{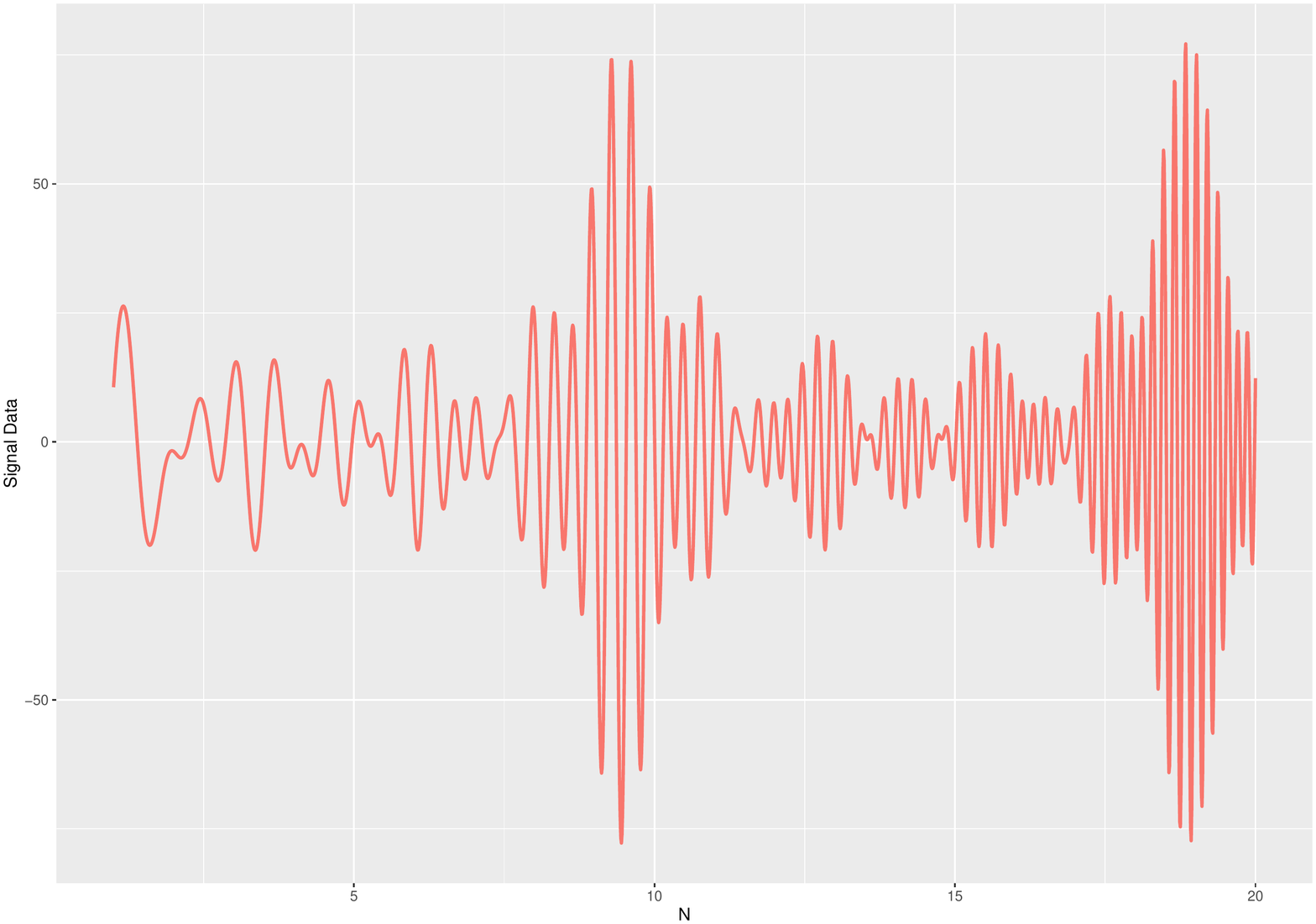}}}\\
		\caption{{\textit{Chirp waveform of a multi-component chirp model with equal chirp rates.}}}
		\label{chirp_wav}
	\end{figure}

	In this paper, we have considered the following real valued multi-component chirp model with equal chirp rates parameters: \begin{align} \label{true model}
		y(n)=&\nonumber\displaystyle\sum_{k=1}^{p}\bigg(A_k^0\cos(\alpha_k^0 n+\beta^0n^2)+B_k^0\sin(\alpha_k^0 n+\beta^0n^2)\bigg)\\&+X(n),\hspace{10pt} n=1,2,3,\ldots,N ,
	\end{align} 
	where, 	\(\bm{\theta}^0=
	(\bm{A}^{0\top},  \bm{\xi}^{0\top})^\top=\) \\\text{}\hspace{30pt} \((\underbrace{A_1^0,B_1^0,A_2^0,B_2^0,\ldots,A_p^0,B_p^0}_{\bm{A}^\top},\underbrace{\alpha_1^0,\alpha_2^0,\ldots,\alpha_p^0,\beta^0}_{\bm{\xi}^\top})^\top\) is the parameter vector, \(A_k^0, B_k^0\) are the amplitude parameters, \(\alpha_k^0\) is the frequency parameter, \(\beta^0\) is the chirp rate parameter of the \(k^{th}\) component, \(k=1,2,3,\ldots,p\) and \(X(n)\) is a noise term.  {We present Fig.\ref{chirp_wav} for illustration of  chirp waveform of the data generated from model \eqref{true model} for two different values of \(p = 5, 10\)}.
	\\~\\
	{Consider the problem of radar tracking of moving targets, where translational motion of the target has been compensated,  it has been observed  in \cite{Rao_2014}  that the Doppler phase of the signal reflected from the scatterer  to be a quadratic function of time. Then, the
		cross-range signal in each range bin after slant range migration correction (see \cite{Rao_2014}) can be regarded as multi-component chirp model with a common unknown chirp rate. Another application where the model \eqref{true model} emerges is, when a chirp signal is transmitted from a radar or a communication
		system (e.g., linear frequency modulated signal (LFM))
		and passes through a linear finite impulse response (FIR) channel (e.g., with multipath
		propagation), a multi-component chirp model emerges at the
		output with each component having identical highest order
		coefficient,
		which means that the filter output
		is composed of multi-component chirp model sharing the same highest
		order coefficient, i.e., chirp rate for quadratic phase, see  \cite{Barbar_1998} for more details.
		Multi-component chirp model with equal chirp  rates also appear in studying high resolution radar imaging of the Earth from a satellite or an aircraft (\cite{Barbar_1998}).
		It has been shown through analysis in \cite{Aifang_2003},  that the
		echo signal from each scatterer of fast-moving target is
		approximately a multi-component chirp signal with the equal chirp rates. It has also been concluded in \cite{Aifang_2005}  that  the ISAR received signal from the moving target,
		which has many scatterers, is approximately a multi-component chirp signal with the equal chirp rates.
		Similar applications can be observed for an uncooperative target
		with multi-scatterers, the echo signals after heterodyne
		detection can be considered as multi-component LFM signals
		with the equal chirp rates,\cite{Ruan_2013}. Some other applications mentioned in the literature are, 	the echo data in fast-time dimension has the form of multi-component chirp signals echo  of high speed targets, \cite{Tian_2013}, \cite{Wu_2014} etc. Therefore, model \eqref{true model} has vast applications in radar and engineering domains and hence studying efficient estimation problem would be of considerable interest to the aerospace and engineering audience.	
		References of such models may be found in \cite{Barbar_1995}, \cite{Ma_2006} and \cite{Ma_2004}.} \\~\\
	Substantial literature is available on parameter estimation of multi-component chirp models with distinct chirp rates  (e.g., see \cite{Wang_2006}, \cite{Wang_2008}, \cite{Yang_2015},  \cite{Lahiri_2015}, \cite{Nandi_2021}, \cite{Xu_2018} and the references cited therein). On the other hand, a few works in the literature are dedicated for multi-component chirp models with equal chirp rates (e.g., see \cite{Barbar_1998} and \cite{Tian_2013}).
	LSEs would be the most natural choice to obtain optimal estimators of parameters of model \eqref{true model}.  However, computing LSEs would require solving a $(p+1)$ dimensional non-linear optimization problem, which will be  computationally prohibitive in practice. Hence, there is a need to develop  computationally efficient methods with performance  matching with the optimal LSEs.
	The model \eqref{true model} is a sub-model of multi-component chirp signal with distinct  chirp rates. Therefore, one may think of implementing efficient procedures proposed for multi-component chirp signal with distinct chirp rates, for the model \eqref{true model}. One such example is sequential least squares (see \cite{Lahiri_2015}), which is shown to be computationally simpler than LSEs, and also have same asymptotic distribution  as that of LSEs. However, since chirp rates \((\beta^0)\) of different components are same for \eqref{true model},
	it loses the orthogonality structure, and hence  it becomes
	a challenging and non-trivial problem to find a computationally efficient estimation procedure. Therefore, it is not immediate how the sequential estimators will behave both theoretically as well as empirically.\\~\\
	Sub-optimal estimators like  product high order ambiguity function (PHAF) and integrated cubic phase function (ICPF) of \cite{Wang_2008} are implemented in \cite{Barbar_1998} and \cite{Tian_2013}, respectively, for the model \eqref{true model}. PHAF provides sub-optimal estimator while ICPF based estimators are computationally burdensome. However, theoretical analysis of PHAF and ICPF based estimators is quite challenging for a multi-component model, because of the  interferences and/or interactions among different components and noise. These interferences  occur due to non-linear operations  of these methods on observed data (see \cite{Barbar_1998} and \cite{Wang_2008}).  Also, de-chirping is performed to obtain estimators of lower order coefficients parameters. This  severely reduces the performance of estimators of lower order coefficients, caused by the error propagation from highest order coefficient estimates.
	The aim of our paper is two-fold:
	\begin{itemize}
		\item 	derive theoretical properties of least squares estimators for the model parameters of \eqref{true model};
		\item propose computationally and asymptotically efficient estimators of the chirp rate parameters, as an alternative to LSEs.
	\end{itemize}
	{We have considered a model which is of significant importance in radar applications of tracking mobile target. Our proposed estimation methods are motivated from the idea of reducing computational complexity of least squares estimators while, maintaining the optimal rates of convergence of the estimators. Combined estimator is based on the principle of improving estimators of chirp rate based on sequential process, while, plugin estimator is based on minimizing the computational load of LSEs to the least.   Technical significance of our proposed estimators lies in obtaining same rates of convergence as that of LSEs along with the benefit of reducing computational complexity.
		We have also shown that combined  estimator of chirp rate parameter to be asymptotically optimal. We have further analysed a radar data available in \cite{ISAR_matlab_book} and shown that our proposed estimators can be efficiently used to reconstruct the ISAR image of a noisy radar data.}
	
	The rest of the paper is organised as follows: we first describe the assumptions for motivating the estimation methodology of all the three proposed estimators in Section \ref{est_methd}.  We also provide detailed estimation methodology and theoretical results of the estimators  in the same Section \ref{est_methd}.   Detailed numerical simulations are  executed in Section \ref{numerical} to study the performance of proposed estimators and also comparing the performance with ICPF, in terms of mean squared errors (MSEs) and computational time. We have also proposed some implementation procedures using PHAF based estimators, to overcome the computational challenge with the two-dimensional (2D) optimization problem associated with proposed estimators, along with its careful assessment on one radar data.  The paper is concluded in Section \ref{concl} followed by the appendices containing proofs of the theorems.     
	\section{Estimation Methodology} \label{est_methd}
	We first mention the assumptions on   noise term and parameters of the model \eqref{true model}, required to motivate the layout of the proposed methodology, as well as to derive the asymptotic theoretical properties of all the three estimators.
	\subsection{Assumptions}
	\begin{assump}\label{assump_1}
		$X(n)$ is stationary linear process with the following form: 
		\begin{equation} \label{err_1}
			X(n) = \displaystyle{\sum_{j=-\infty}^{\infty}}a(j)\epsilon(n-j), \end{equation}  
		such that the coefficients \(a(j)\) are absolutely summable, i.e., 
		\begin{equation} \label{err_2}
			\hspace{5pt}\displaystyle{\sum_{j=-\infty}^{\infty}}|a(j)|<\infty, \end{equation} 
		and $\{\epsilon(n)\}$ is a sequence of independently and identically distributed (i.i.d.) random variables with mean $0$, variance $\sigma^2$ and  finite fourth moment.
	\end{assump}
	\begin{assump}\label{assump_2}
		True value of parameters \(A_k^0,B_k^0\in [-M,M]\) for some \(M>0\), \(\alpha_k^0\in (0,2\pi)\), \( \beta^0\in(0,\pi/2)\) and $A_k^{0^2}+B_k^{0^2} > 0$ for \(k=1,2,3,\ldots,p\).
	\end{assump}
	\begin{assump}\label{assump_3}
		$A_1^{0^2}+B_1^{0^2} > A_2^{0^2}+B_2^{0^2}>\cdots >A_p^{0^2}+B_p^{0^2}. $
	\end{assump}
	Assumption \ref{assump_1} defines a large class of errors in the sense that it contains  the usual assumption of i.i.d. Gaussian errors as well as  moving average processes, autoregressive processes etc., and hence, including the dependent structure too. Assumption \ref{assump_2} is a standard assumption, which mainly indicates that the frequencies and frequency rate parameters are not on the boundary{, and at the same time, it avoids any identifiabiliity problem in possible values of the parameter space}.	Assumption \ref{assump_3} helps in the motivation of developing  a natural sequential procedure and also guarantees the consistency of the estimators obtained through this sequential process.  Assumption \ref{assump_3} also guides sequential procedure in identifying the different components in correct order. \\~\\
	Next, we describe the estimation methodologies for LSEs, sequential combined estimators and  sequential plugin estimators of the parameters of model  \eqref{true model}. These methods are based on the assumption that  the number of components \(p\) is known apriori, for the model \eqref{true model}.
	\subsection{LSEs}\label{LSEs_methd}
	Let us consider \(\bm{Y}=(y(1),y(2),\ldots,y(N))^\top\) to be the data vector and 	define the following sum of squares: \[Q(\bm{\theta})=(\bm{Y}-\bm{W}(\bm{\xi})\bm{A})^\top(\bm{Y}-\bm{W}(\bm{\xi})\bm{A}),\]
	where, \(\bm{\theta}=
	(\bm{A}^{\top},  \bm{\xi}^{\top})^\top\), \((n,2k-1)^{th}\) and \((n,2k)^{th}\) elements of the \(N\times 2p\) order matrix \(\bm{W}(\bm{\xi})\) are {
		\(
		\cos(\alpha_{k} n+\beta n^2)\) and \(\sin(\alpha_{k} n+\beta n^2)\) }respectively.
	Then, LSEs of \(\bm{\theta}^0\) will be obtained as follows:
	\begin{equation}\label{lses_expr}
		\widehat{\bm{\theta}} =\arg\min\limits_{\bm{\theta}}Q(\bm{\theta}).
	\end{equation} 
	Note that using separable linear regression technique, above optimization problem can be reduced to a \((p+1)\) dimensional from a \((3p+1)\) dimensional  non-linear optimization problem (see e.g., \cite{Lahiri_2015} and  \cite{Richards_1961}). {So, after simplification, above estimators can be obtained equivalently as \[\widehat{ \bm{\xi}} = \arg\min\limits_{\bm{\theta}}\big(\bm{Y}-\bm{W}(\bm{\xi})\bm{A}(\bm{\xi})\big)^\top\big(\bm{Y}-\bm{W}(\bm{\xi})\bm{A}(\bm{\xi})\big) ,\]
		where, \(\bm{A}(\bm{\xi}) = \Big(\bm{W}(\bm{\xi})^\top\bm{W}(\bm{\xi})\Big)^{-1}\bm{W}(\bm{\xi})^\top\bm{Y},\) for more details, one can refer \cite{Grover_2021}. Here, estimate of \(\bm{A}\) is given by:\[
		\widehat{\bm{A}} =\Big(\bm{W}(\widehat{\bm{\xi}})^\top\bm{W}(\widehat{\bm{\xi}})\Big)^{-1}\bm{W}(\widehat{\bm{\xi}})^\top\bm{Y} .\]}
	\subsection{Sequential Combined Estimators}\label{seq_comb_methd}
	We use the idea of sequential estimation procedure  \cite{Lahiri_2015} to find computationally simpler methods. This sequential estimator is based on  estimating only one component at a time, hence requires only a 2D optimization problem to solve \(p\) number of times.  We now define the sequential combined estimators of the parameters of model \eqref{true model} in following steps:
	\begin{itemize}
		\item \textbf{Step-1} 	Consider the following sum of squares:	\begin{align*} Q_1(\bm{\theta_1})&=\displaystyle \sum_{n=1}^{N}\big(y(n) -A_1\cos(\alpha_1 n+\beta_1 n^2)\\&-B_1\sin(\alpha_1 n+\beta_1 n^2)\big)^2.\end{align*} 
		First component parameter vector is \(\bm{\theta}_1^0=({A}_1^0,{B}_1^0,\alpha_1^0,\beta^0)^\top\) and let $\bm{\theta}_1=({A}_1,{B}_1,\alpha_1,\beta_1)^\top$, then we define sequential combined estimator of  the first component parameters as follows: 
		\begin{equation}\label{first_comb_ssq}
			\breve{\bm{\theta}}_1 =\arg\min\limits_{\bm{\theta}_1}Q_1(\bm{\theta}_1).
		\end{equation} 
		Note that above \eqref{first_comb_ssq} can be reduced to a 2D optimization problem by using separable linear regression technique (see \cite{Richards_1961}).{ We can simplify \eqref{first_comb_ssq} similar to that of \eqref{lses_expr}.}
		It will also  be  a 2D optimization problem for the  remaining component estimators.
		\item \textbf{Step-2} We eliminate the effect of first component from the data and then define sequential combined estimator $\breve{\bm{\theta}}_2=(\breve{A}_2,\breve{B}_2,\breve{\alpha}_2,\breve{\beta}_2)^\top$  of second component parameter vector  \(\bm{\theta}_2^0=({A}_2^0,{B}_2^0,\alpha_2^0,\beta^0)^\top\)  by updating data as:
		\[\breve{y_1}(n)=	y(n)-\breve{A}_1\cos(\breve{\alpha}_1 n+\breve{\beta}_1 n^2)-\breve{B}_1\sin(\breve{\alpha}_1 n+\breve{\beta}_1 n^2).\mbox{ }\] Consider, 
		\begin{align*} 
			Q_2(\bm{\theta}_2)&=\displaystyle \sum_{n=1}^{N}\big(\breve{y_1}(n) -A_2\cos(\alpha_2 n+{\beta}_2 n^2)\\&-B_2\sin(\alpha_2 n+{\beta}_2 n^2)\big)^2,
		\end{align*} 
		where $\bm{\theta}_2=({A}_2,{B}_2,\alpha_2,\beta_2)^\top$, then the second component estimator is given by the following equation:
		\begin{equation}\label{second_comp_combin}
			\breve{\bm{\theta}}_2 =\arg\min\limits_{\bm{\theta}_2}Q_2(\bm{\theta}_2).
		\end{equation} 
		{ We can simplify \eqref{second_comp_combin} similar to that of  \eqref{lses_expr}.}
		\item \textbf{Step-3} We  repeat step-2  for the remaining components' sequential combined estimators, i.e. $\breve{\bm{\theta}}_k=(\breve{A}_k,\breve{B}_k,\breve{\alpha}_k,\breve{\beta}_k)^\top$, \(k\geq 3.\) 
		\item \textbf{Step-4} 
		Since all of the different components in model \eqref{true model} have equal chirp rates parameters, so we propose combine estimators of \(\beta^0\) from each component in an optimal manner.	
		It is known that, when the chirp rates are different, then estimators of distinct components are asymptotically independent, (see \cite{Lahiri_2015}). Therefore, using this result as motivation, we  propose sequential combined   estimator \(\breve{\beta}\) of $\beta^0$  by minimizing the variance of linear combination of all component estimators \(\breve{\beta}_k\) among the following class:
		$$\breve{\beta}=\displaystyle \sum_{k=1}^{p}l_k\breve{\beta}_k,\hspace{20pt}\displaystyle \sum_{k=1}^{p}l_k=1.$$
		Sequential combined estimator of chirp rate parameter \(\beta^0\) is therefore,  given by:\[\breve{\beta}=\cfrac{1}{\displaystyle \sum_{k=1}^{p}(\breve{A}_k^{0^2}+\breve{B}_k^{0^2})}\displaystyle \sum_{k=1}^{p}(\breve{A}_k^{0^2}+\breve{B}_k^{0^2})\breve{\beta}_k.\] 
	\end{itemize}	
	We have shown that \(\breve{\beta}\) is an asymptotically optimal estimator of chirp rate parameter \(\beta^0\). Finally, \(\breve{\bm{\theta}}=(\breve{A}_1,\breve{B}_1,\breve{A}_2,\breve{B}_2,\ldots,\breve{A}_p,\breve{B}_p,\breve{\alpha}_1,\breve{\alpha}_2,\ldots,\breve{\alpha}_p,\breve{\beta})^\top\) is the proposed sequential combined estimator of parameter \(\bm{\theta}^0=(A_1^0,B_1^0,A_2^0,B_2^0,\ldots,A_p^0,B_p^0,\alpha_1^0,\alpha_2^0,\ldots,\alpha_p^0,\beta^0)^\top\).
	\subsection{Sequential Plugin Estimators}\label{seq_plug_methd}
	It is possible that the number of components in model \eqref{true model} may be very large for many real-life applications.  Sequential combined estimators will require \(p\) number of 2D optimization problems to solve. In this situation, implementing sequential combined estimators  will involve a tedious computation. So, to resolve this issue, we perform 2D optimization only for the first component estimators and then, we plugin this estimate of \(\beta^0\) in all remaining components. Hence the sequential plugin estimator will require  one 2D and \((p-1)\) number of 1D optimization problems to solve, reducing the computational cost significantly. We now describe the detailed methodology of  sequential plugin estimators of parameters of model \eqref{true model} in following steps:
	\begin{itemize}
		\item \textbf{Step-I} 	Define following sum of squares to estimate first component parameters of model \eqref{true model}:	\begin{align*} Q_1(\bar{\bm{\theta}}_1)&=\displaystyle \sum_{n=1}^{N}\big(y(n) -A_1\cos(\alpha_1 n+\beta n^2)\\&-B_1\sin(\alpha_1 n+\beta n^2)\big)^2.
		\end{align*} 
		Further define $\bar{\bm{\theta}}_1=({A}_1,{B}_1,\alpha_1,\beta)^\top$, then we estimate the first component parameters \(\bar{\bm{\theta}}_1^0=({A}_1^0,{B}_1^0,\alpha_1^0,\beta^0)^\top\) as follows: 
		\begin{equation}\label{first_plug_in_ssq}
			\widetilde{\bm{\theta}}_1 =\arg\min\limits_{\bar{\bm{\theta}}_1}Q_1(\bar{\bm{\theta}}_1).
		\end{equation} 
		{ We can simplify \eqref{first_plug_in_ssq} similar to that of  \eqref{lses_expr}.}
		Equation \eqref{first_plug_in_ssq} can be reduced to a 2D optimization problem. But for other components, it will be a 1D optimization problem (see \eqref{secnd_plug_in_ssq}). It can be  observed that sequential combined estimators and sequential plugin estimators of the first component parameters are same.
		\item \textbf{Step-II} To compute second component estimators $	\widetilde{\bm{\theta}}_2=( \widetilde{A}_2,\widetilde{B}_2,\widetilde{\alpha}_2)^\top$ of \(\bar{\bm{\theta}}_2^0=(A_2^0,B_2^0,\alpha_2^0)^\top\),  we update the data as:
		\[\widetilde{y_1}(n)=	y(n)-\widetilde{A}_1\cos(\widetilde{\alpha}_1 n+\widetilde{\beta} n^2)-\widetilde{B}_1\sin(\widetilde{\alpha}_1 n+\widetilde{\beta} n^2),\]
		and define following sum of squares: \[Q_2(\bar{\bm{\theta}}_2)=\displaystyle \sum_{n=1}^{N}\big(\widetilde{y_1}(n) -A_2\cos(\alpha_2 n+\widetilde{\beta} n^2)-B_2\sin(\alpha_2 n+\widetilde{\beta} n^2)\big)^2,\]  where  \(\bar{\bm{\theta}}_2=(A_2,B_2,\alpha_2)^\top\).
		Second component sequential combined estimators are given by:
		\begin{equation}\label{secnd_plug_in_ssq}
			\widetilde{\bm{\theta}}_2 =\arg\min\limits_{\bar{\bm{\theta}}_2}Q_2(\bar{\bm{\theta}}_2).
		\end{equation} 
		\item \textbf{Step-III} Repeat step-II for remaining component parameters i.e., \(\bar{\bm{\theta}}_k^0=(A_k^0,B_k^0,\alpha_k^0)\) for \(k\geq 3\).
	\end{itemize}
	In this manner, we obtain sequential plugin estimator  \(\widetilde{\bm{\theta}}=(\widetilde{A}_1,\widetilde{B}_1,\widetilde{A}_2,\widetilde{B}_2,\ldots,\widetilde{A}_p,\widetilde{B}_p,\widetilde{\alpha}_1,\widetilde{\alpha}_2,\ldots,\widetilde{\alpha}_p,\widetilde{\beta})^\top\)  of the parameter\\ \(\bm{\theta}^0=(A_1^0,B_1^0,A_2^0,B_2^0,\ldots,A_p^0,B_p^0,\alpha_1^0,\alpha_2^0,\ldots,\alpha_p^0,\beta^0)^\top\).
	We emphasize that \(\beta^0\) is estimated only once in sequential plugin method as compared to the case of sequential combined, where \(\beta^0\) is estimated repeatedly for all different component parameters.
	
	\subsection{Main Theorems}\label{thm_main}
	\begin{theorem}\label{cons_tm}
		Estimators \(\widehat{\bm{\theta}},\)  \(\breve{\bm{\theta}}\) and \(\widetilde{\bm{\theta}}\) are strongly consistent for the parameter \(\bm{\theta}^0\) of model \eqref{true model} under the Assumption \ref{assump_1}, \ref{assump_2} and \ref{assump_3}, i.e., \(\widehat{\bm{\theta}}\xrightarrow{a.s.}\bm{\theta}^0\), \(\breve{\bm{\theta}}\xrightarrow{a.s.}\bm{\theta}^0\), \(\widetilde{\bm{\theta}}\xrightarrow{a.s.}\bm{\theta}^0\) as \(N\rightarrow\infty\). Here, \(a.s.\) represents almost sure convergence.
	\end{theorem}
	\begin{proof}
		Please refer \nameref{consis_proof} for the proof.
	\end{proof}
	\begin{theorem}\label{asymp_norm_thm}
		Estimators \(\widehat{\bm{\theta}},\)  \(\breve{\bm{\theta}}\) and \(\widetilde{\bm{\theta}}\) follow asymptotically normal distributions under the Assumption \ref{assump_1}, \ref{assump_2} and \ref{assump_3}, with the scaling matrix\\ \begin{small} \(\bm{D}^{-1}=diag(\underbrace{N^{1/2},N^{1/2},\ldots,N^{1/2},N^{1/2}}_{2p\hspace{5pt}\mbox{times}}, \underbrace{N^{3/2},\ldots,N^{3/2}}_{p \hspace{5pt} \mbox{times}} ,N^{5/2})\)\end{small} 
		, i.e., 
		\[\bm{D}^{-1}\big(\widehat{\bm{\theta}}-\bm{\theta}^0\big)\xrightarrow{d}\mathcal{ N}_{3p+1}\Big(\bm{0},2c\sigma^2\bm{\Sigma}_1\Big),\]
		\[\bm{D}^{-1}\big(\breve{\bm{\theta}}-\bm{\theta}^0\big)\xrightarrow{d}\mathcal{ N}_{3p+1}\Big(\bm{0},2c\sigma^2\bm{\Sigma}_2\Big) \mbox{ and} \]
		\[\bm{D}^{-1}\big(\widetilde{\bm{\theta}}-\bm{\theta}^0\big)\xrightarrow{d}\mathcal{ N}_{3p+1}\Big(\bm{0},2c\sigma^2\bm{\Sigma}_3\Big),\]
		where \(diag(a_1,a_2,\ldots,a_k)\) represents \(k\times k\) diagonal matrix with diagonal entries \(a_1,a_2,\ldots,a_k\). Diagonal elements of positive definite matrices \(\bm{\Sigma}_1,\bm{\Sigma}_2\) and  \(\bm{\Sigma}_3\) are described in equations \eqref{asymp_lses_bet_alph}-\eqref{last_eqn} of \nameref{normal_proof}.
	\end{theorem}
	\begin{proof}
		Please refer \nameref{normal_proof} for the proof.
	\end{proof}
	{ Results similar to Theorem \ref{cons_tm} and \ref{asymp_norm_thm} exist for the estimators already exist for the estimators of parameters of multi-component chirp model with distinct chirp rates. However, the estimators of multi-component chirp model with equal chirp rates needed separate attention because in this case, the usual asymptotic orthogonality of different chirp components is lost. The asymptotic variance-covariance matrix of the estimators for equal chirp rates is different from the distinct chirp rates case. So, the novelty of our work lies in deriving separately all the properties of the estimators, which do not follow trivially from the properties of estimators developed for distinct chirp rates case.}\\~\\
	\noindent\textbf{Remark:}	If \(X(n)\) is an i.i.d. sequence following normal distribution with mean 0 and variance \(\sigma^2\) in Theorem \ref{asymp_norm_thm}, then diagonal entries of \(\bm{\Sigma}_1\) will provide the corresponding Cramer-Rao lower bound. We have reported asymptotic variance of LSEs of \(\alpha_k^0\) and \(\beta^0\) in \eqref{asymp_lses_bet_alph} and \eqref{asymp_lses_bet_alph2}. It has been shown that sequential combined estimators of chirp rate parameter \(\beta^0\) is asymptotically optimal, however, the asymptotic variance of sequential plugin estimator of frequency parameter \(\alpha_k^0\) is less than that of sequential combined estimator (see  comparison in page 3 of the supplementary material). Therefore, \(\bm{\Sigma}_3-\bm{\Sigma}_2\) is neither positive definite, nor negative definite. These theoretical results are also validated by numerical simulations in the next section.  
	\section{Numerical Simulations}\label{numerical}
	We have performed extensive numerical simulations to study the behaviour of the different proposed methods based on their MSEs, and also about the time  needed to obtain these estimators. Numerical simulations have been categorized broadly into five subsections. We compare the finite sample performance of LSEs, sequential plugin estimators and sequential combined estimators of the non-linear parameters,  in terms of  MSEs, for a five-component chirp model with equal chirp rates in the subsection \ref{perform_three}. We have compared our proposed estimators with ICPF based estimator  of chirp rate parameter in subsection \ref{icpf_compar}.   Then, we have presented computational cost in terms of time to obtain these estimators in subsection \ref{time}. We also provide some implementations and  performance of the estimators in subsection \ref{implement}  to reduce computational problem for the proposed estimators.  Sub-section \ref{radar_data} includes assessment of the proposed implementation procedure using one radar signal data.
	Throughout the numerical simulations, we have used  analytical form of the model \eqref{true model} for comparison, refer equation (1) of \cite{Barbar_1998}. We have also considered the following two different kinds of error sequence \(X(n)\) as:
	\begin{itemize}
		\item \textbf{E1:} \(X(n)\) is a sequence of  independently and identically distributed (i.i.d.) Gaussian errors with mean 0 and variance $\sigma^2$;
		\item \textbf{E2:}  \(X(n)\) is autoregressive moving average (ARMA) errors with following representation:
		\begin{align}\label{err_cas_2}
			X(n) =& 0.6X(n-1)+\epsilon(n)+0.1\epsilon(n-1).
		\end{align}
		where $\epsilon(n)$ is a sequence of i.i.d. Gaussian random variables with mean 0 and variance $\sigma^2$.  
	\end{itemize}	The real and imaginary parts are contaminated independently with the above two kind of errors E1 and E2.
	Parameters of the model for the subsections \ref{perform_three}, \ref{icpf_compar} and \ref{time} are as follows:
	\begin{align} \label{par_set}
		&p=5, \hspace{5pt}A_1^0=3.35, \hspace{5pt}A_2^0=2.8, \hspace{5pt}A_3^0=2.1, \hspace{5pt} A_4^0=1.59,\nonumber \\& \hspace{5pt}A_5^0=0.9, \hspace{5pt} \nonumber\alpha_1^0=0.89, \hspace{5pt}\alpha_2^0=0.96, \hspace{5pt}\alpha_3^0=0.76, \\  &\hspace{5pt} \alpha_4^0=0.56, \hspace{5pt}\alpha_5^0=0.37, \hspace{5pt}\beta^0=0.87 .
	\end{align}
	\subsection{Comparing MSEs of LSEs with sequential  combined  and sequential plugin estimators}\label{perform_three}
\begin{figure}
		\centering
		\subfloat[\textit{Estimators of $\alpha_1^0$,  $\alpha_2^0$ and $\alpha_3^0$. }
		]{\label{fig:alp1}{\includegraphics[width=0.9\linewidth]{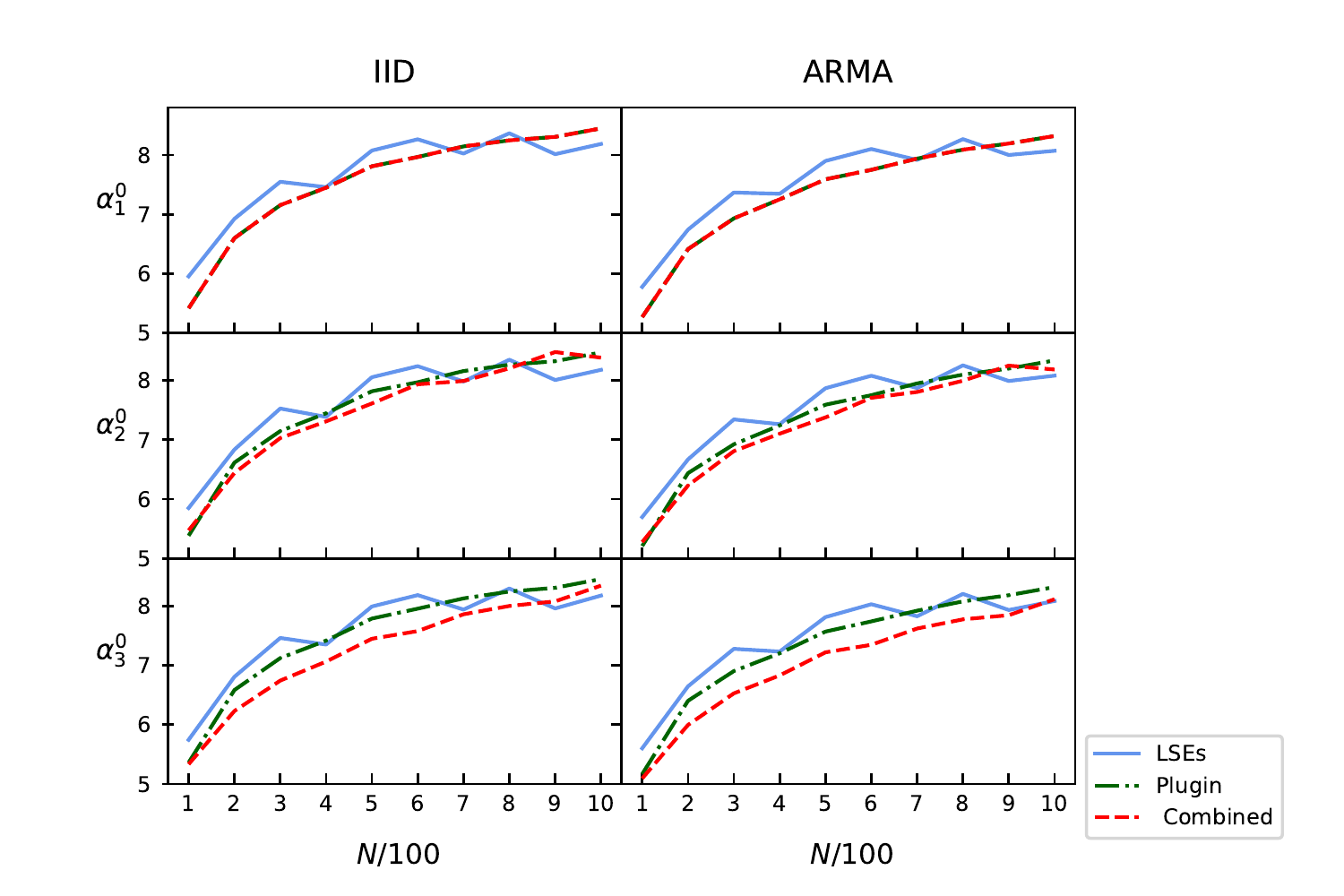}}}\qquad
		\subfloat[\textit{Estimators of $\alpha_4^0$, $\beta^0$ and $\alpha_5^0$.} ]{\label{fig:beta1}{\includegraphics[width=0.9\linewidth]{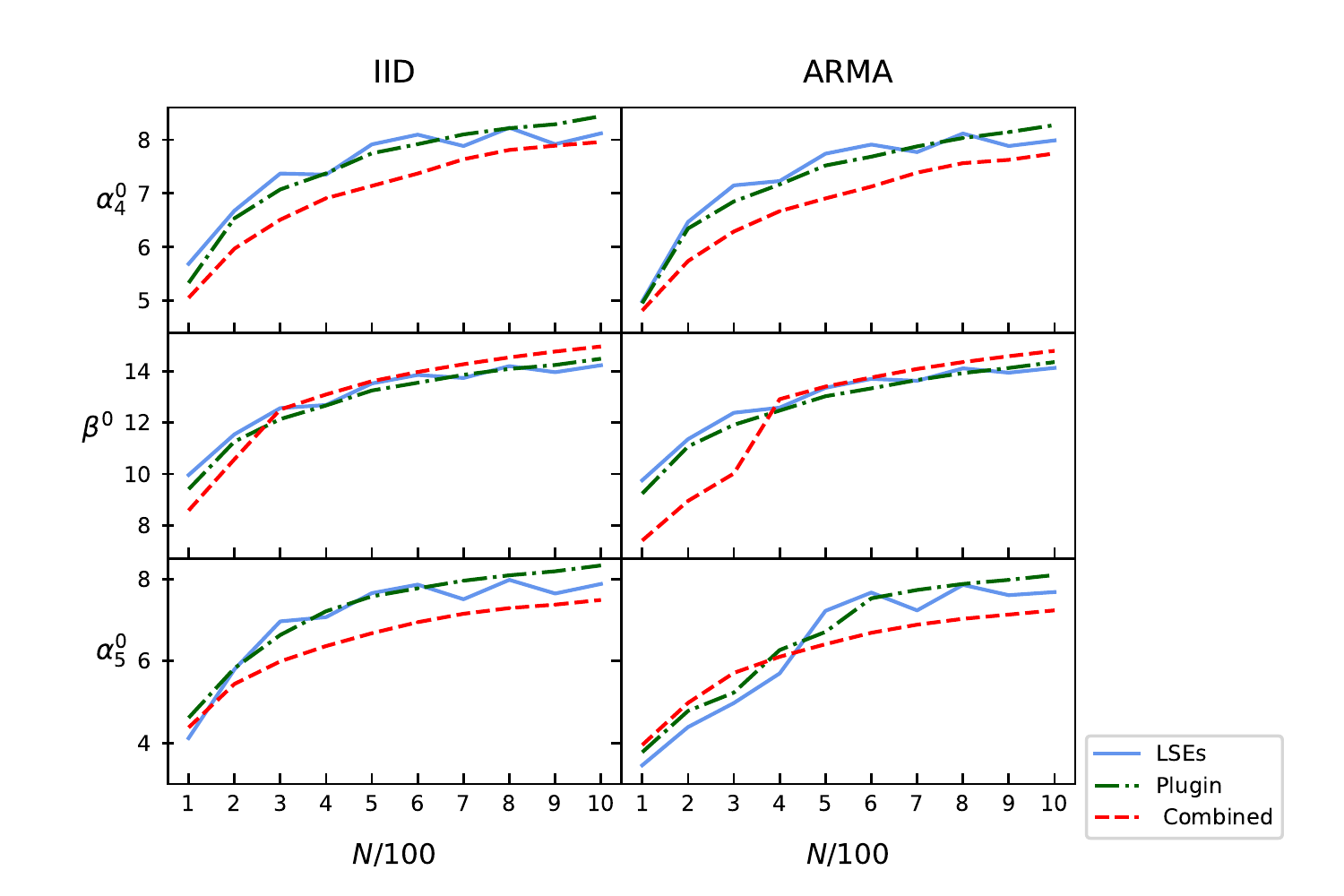}}}\\
		\caption{\textit{MSEs (on \(-\log_{10}\) scale) of LSEs, sequential plugin estimators  (Plugin) and sequential combined estimators (Combined)   versus  the increasing sample size, for the case \(\sigma=2\)}.}
		\label{plot_first_comp}
	\end{figure}

\begin{figure}
		\centering
		\subfloat[\textit{Estimators of $\alpha_1^0$,  $\alpha_2^0$ and  $\alpha_3^0$. }
		]{\label{fig:alp3}{\includegraphics[width=0.92\linewidth]{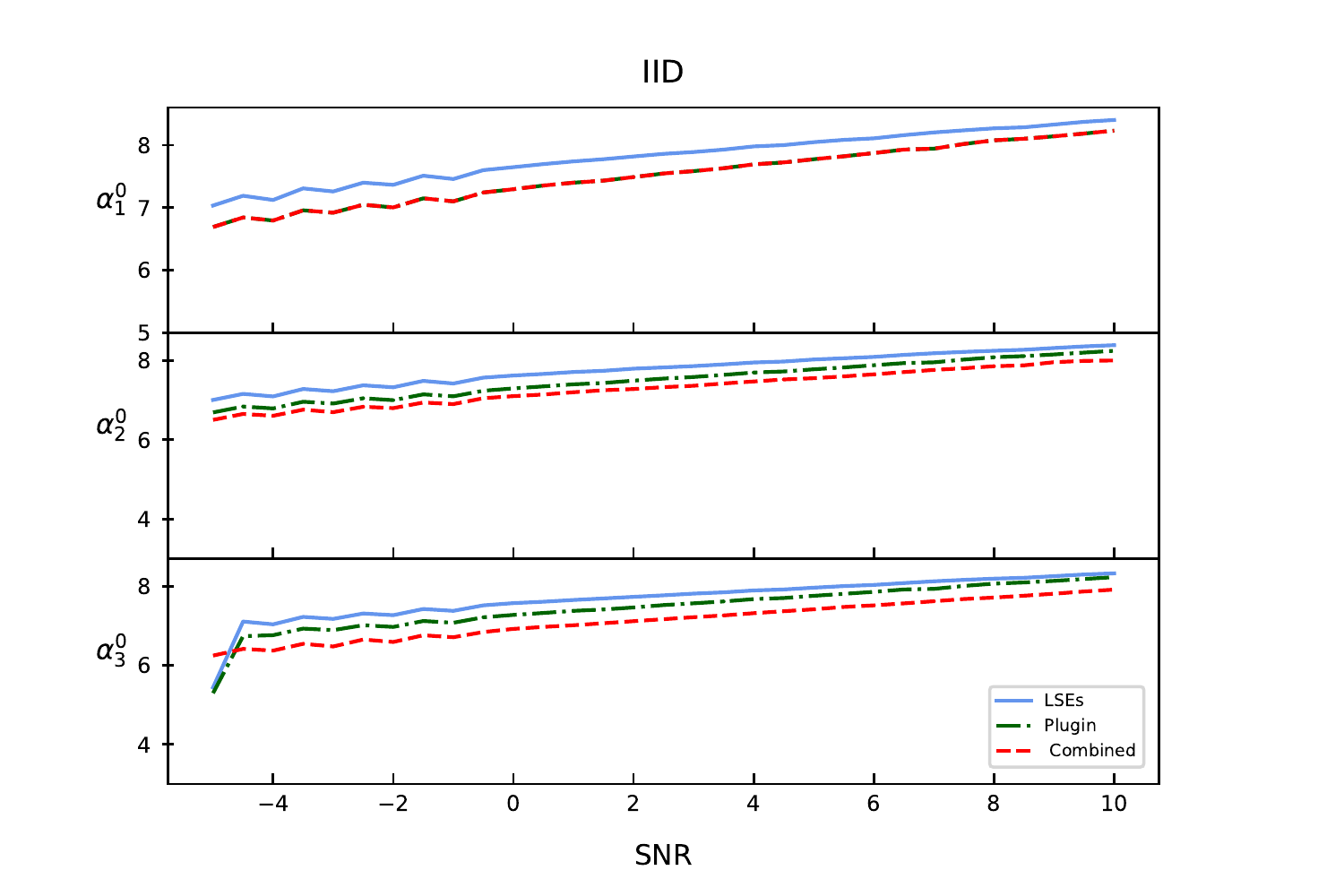}}}\qquad
		\subfloat[\textit{Estimators of $\alpha_4^0$, $\beta^0$ and $\alpha_5^0$.}]{\label{fig:beta3}{\includegraphics[width=0.92\linewidth]{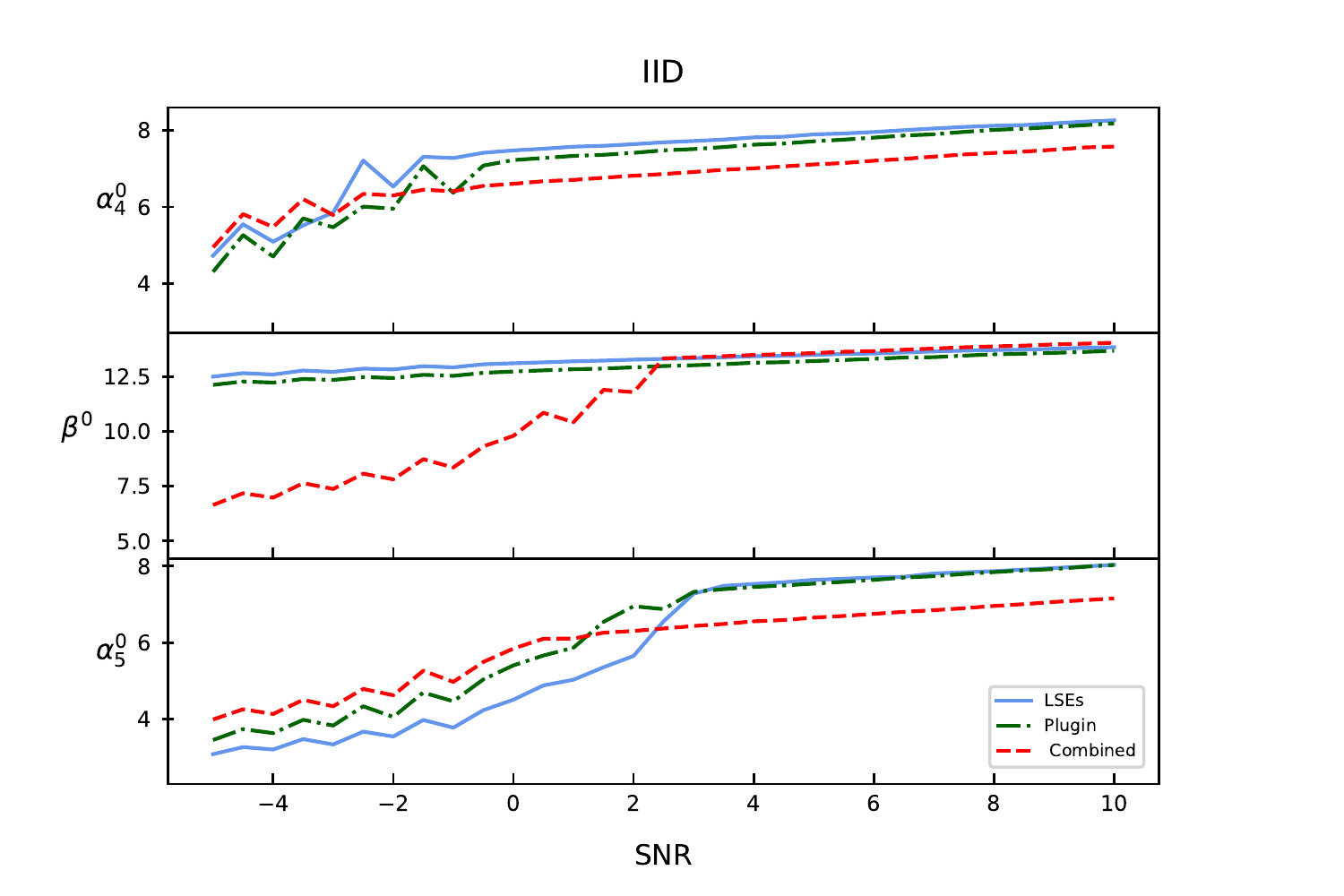}}}\\
		\caption{\textit{MSEs (on \(-\log_{10}\) scale) of LSEs, sequential plugin estimators  (Plugin) and sequential combined estimators (Combined)   versus the SNR, for fixed sample size \(N=500\)}.}
		\label{plot_first_comp3}
	\end{figure}

	For the analytical model (1) of \cite{Barbar_1998}, we have studied finite sample performance of LSEs, sequential combined estimators and sequential plugin estimators in the sense of MSEs. We have taken different sets of sample sizes \(N=100,200,\ldots,1000\).  Both E1 and E2 are considered for error \(X(n)\) with \(\sigma=2,3\). For a generated data-set, we have obtained estimates of LSEs, sequential  combined,  and sequential plugin estimators estimators. All of these estimators do not have any explicit form, so we have to resort to some standard  numerical procedures. Here, we have implemented  the Nelder–Mead simplex algorithm \cite{Neld_1965} ( using ``optim"  function in R software) for optimization of the associated objective function to obtain all these estimators. We also need good initial guess for convergence of the Nelder-Mead algorithm to the true parameters. So, we have used grid-search on the intervals \([\alpha_k^0-1/N,\alpha_k^0+1/N]\) for \(k=1,2,3,4,5\) and \([\beta^0-1/N^2,\beta^0+1/N^2]\) to obtain initial guess. 
	\\~\\ Now, we calculate squared errors  between the estimates and true values of the parameters. We then repeat this upto 10,000 number of times and find the average value of squared errors  corresponding to each estimators, which yields MSEs. Plots of MSEs with respect to different sample sizes are provided in Fig. \ref{plot_first_comp} for the case \(\sigma=2\). Similar results obtained for \(\sigma=3\)  are provided  in page 1,  Figure 1 of the supplementary material.\\~\\
	Sub-optimal estimators usually suffer from a relatively higher signal to noise ratio (SNR)
	threshold. Therefore, we have also evaluated MSEs of estimators over 10,000 replications, with respect to 31 different values of SNR ranging from -5 to 10. Errors are sampled randomly from  normal distribution with mean 0 and variance according to the value of SNR. Obtained plot is presented in Fig. \ref{plot_first_comp3}. 
	A list of observations from these figures can be summarised as follows:
	\begin{itemize}
		\item MSEs of all the three estimators decrease as sample size increases, hence validating their consistency as stated in Theorem \ref{cons_tm}  {(Fig. \ref{plot_first_comp})};
		\item MSEs of all the three estimators decrease as SNR increases {(Fig. \ref{plot_first_comp3});
			\item If we consider chirp rate parameter \(\beta^0\), the sequential plugin estimator has more stable MSEs  as compared with that of  sequential combined estimator (Fig. \ref{plot_first_comp})};
		\item SNR threshold for all the three estimators seems to be shifting towards higher value of SNR as the number of components increase, e.g., SNR threshold for all the three estimators of \(\alpha_3^0\) is around -5 dB, whereas it is around 2.5 dB for \(\alpha_5^0\). It also suggests that sequential procedures for models of type \eqref{true model}, perform as good as if we use LSEs in terms of SNR threshold, e.g., see performance of estimators of \(\alpha_5^0\) in Fig. \ref{plot_first_comp3};
		\item MSEs of sequential plugin estimator of the frequency parameters \(\alpha_k^0, k=2,3,4,5\) are comparable  with that of LSEs{ (Fig. \ref{plot_first_comp}, Fig. \ref{plot_first_comp3})};
		\item Sequential combined estimator of chirp rate parameter \(\beta^0\) performs at par with the LSEs for high value of SNRs {(Fig. \ref{plot_first_comp3})};
		\item { Fig. \ref{plot_first_comp} also validates the rate of convergence of obtained estimators, which is provided in Theorem \ref{asymp_norm_thm}. }.
	\end{itemize}
	\subsection{Comparison with ICPF}\label{icpf_compar}
\begin{figure}
		\centering
		\includegraphics[width=0.92\linewidth]{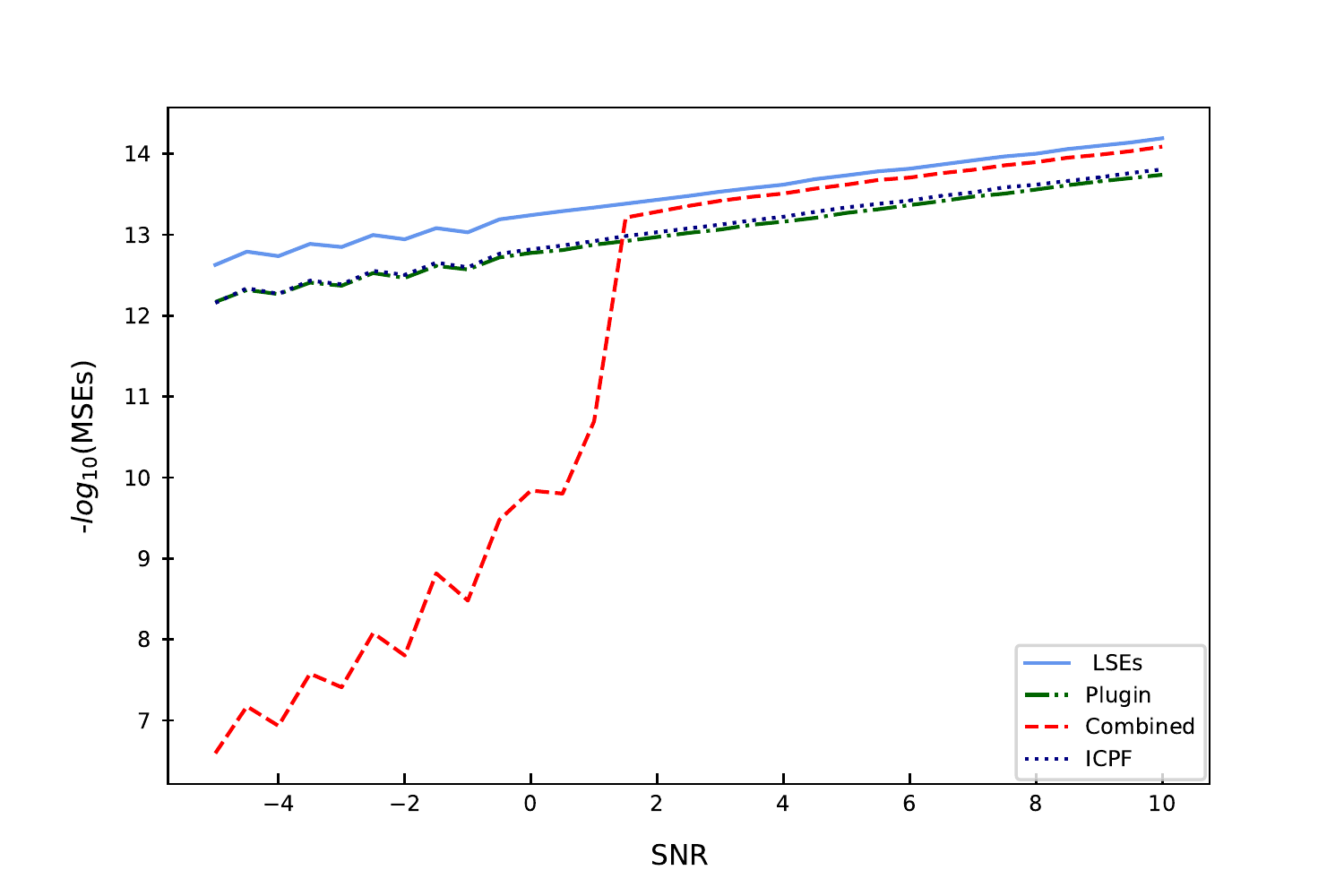}\\
		\caption{\textit{MSEs (on \(-\log_{10}\) scale) of LSEs, sequential plugin estimators (Plugin), sequential combined estimators (Combined) and ICPF based estimator of chirp rate parameter \(\beta^0\)  versus the SNR, for fixed sample size \(N=501\)}.}
		\label{plot_first_comp4}
	\end{figure}

	Implementation of ICPF to estimate chirp rate parameters is shown in \cite{Tian_2013}. We have compared MSEs of LSE,  sequential combined estimator, sequential plugin estimator and ICPF based estimator of chirp rate parameter \(\beta^0\) for the same model considered in  \eqref{par_set}.   We have chosen  30 different values of SNR to compare their performances. MSEs are obtained over 10,000 replications. Simulated results are presented in Fig. \ref{plot_first_comp4}, from which we can conclude the following points:
	\begin{itemize}
		\item 	Sequential combined estimator performs at par with LSEs, and  better than the other computationally efficient estimators, ICPF based estimator and sequential plugin estimator for SNR higher than 2.5 dB;
		\item Sequential plugin estimator has stable MSEs than that of ICPF and sequential combined estimator.
	\end{itemize}  
	Further, we have added Table 1 in supplementary material, for highlighting quantitative technical advantages of the proposed methods.
	\subsection{Computation Time Comparison of all estimators}\label{time}
	We have further presented time cost associated in estimating all  non-linear parameters using LSEs, sequential combined estimators and sequential plugin estimators. We have applied same approach of choosing initial guess and then applying Nelder-Mead to obtain all the final estimates, as suggested in subsection \ref{perform_three}. Figure 2, reported in the page 3 of the supplementary material, shows the ratio of average time (over 1000 replications) taken to compute the LSEs and sequential combined estimator, with respect to the average time to compute sequential plugin estimator. We conclude (Figure 2 of the supplementary material) that the time taken by LSEs is more than 6 times the time taken by the sequential plugin estimators. Also, the time taken by sequential combined estimator is more than  5 times the time taken by sequential plugin estimator for large sample sizes. We had not chosen ICPF in this comparison, because LSEs and sequential combined estimators would normally take much more time for finding good initial guess than ICPF as ICPF requires only 1D optimization problem to solve, whereas LSEs would require 6D optimization problem and sequential combined estimator would require five times 2D optimization problem to solve, for the set-up \eqref{par_set}.\\~\\  However, we have compared the sequential plugin estimator and ICPF estimator on the basis of time required to compute the estimates of \(\beta^0\).  We have used grid-search on the intervals \([\alpha_1^0-1/\sqrt{N},\alpha_1^0+1/\sqrt{N}]\) and \([\beta^0-1/\sqrt{N},\beta^0+1/\sqrt{N}]\) to obtain initial guess of both estimators for implementing Nelder-Mead algorithm. The motivation to use this interval is that we can always use high order ambiguity function (HAF) based estimators to find initial guess of \(\beta^0, \alpha_1^0\) and then perform grid-search on the interval of length  \(1/\sqrt{N}\) around the HAF estimates.
	We have  presented time comparison between our most computationally efficient estimator (sequential plugin estimator) and ICPF based estimator. Here, average time (in seconds) is reported over 50 replications for both estimators. We can see clearly from the Table \ref{time_icpf} that sequential plugin outperforms the ICPF estimator with a significant margin and hence, faster than it.
	Note that we have performed all simulations for time comparison in a system with following specifications: Intel(R) Xeon(R) CPU E7-8870 v3 @ 2.10GHz.
	\begin{table*}
		
		\centering
		\begin{tabular}{|c|c|c|c|c|c|c|c|c|}
			\hline
			Sample Size	&	101	&	121	&	141	&	161	&	181	&	201	&	221	&	241	\\ \hline 
			
			ICPF	&	16.07	&	31.72	&	53.46	&	88.25	&	118.47	&	181.58	&	228.66	&	299.67	\\ \hline
			
			Plugin	&	7.36	&	13.43	&	20.7	&	30.18	&	36.92	&	47.91	&	60.62	&	74.46	\\ \hline

		\end{tabular}
		\caption{\textit{Computational complexity comparison of sequential plugin estimator and ICPF with respect to time (seconds)}.}
		\label{time_icpf}
	\end{table*}

	\subsection{Implementation using PHAF}\label{implement}
	Our most computationally efficient estimator is the sequential plugin estimator, which requires 2D optimization problem to be solved for first component estimators and 1D problem for  remaining components. 2D optimization will be burdensome when sample size is large. So, we now propose an implementation procedure that requires solving only a 1D optimization.
	ICPF and PHAF both require solving a 1D optimization problem, but ICPF needs a large number of algebraic operations to perform on the data and hence making it computationally more burdensome as compared to the PHAF. Therefore, PHAF based estimators can help in reducing computational difficulty associated with obtaining good set of initial guessses for the sequential plugin and sequential combined estimators. We first obtain PHAF estimates of the first component estimators, then we provide following three different options to get initial guess for the first component  sequential plugin estimators as:
	\begin{enumerate}
		\item PHAF;
		\item Use genetic algorithm (GA) in the \(1/(2\pi N)\) neighbourhood of the the PHAF estimates;
		\item Grid search around the \(1/N^{0.9}\)  and   \(1/N^{1.9}\)
		neighbourhood of PHAF estimators of  frequency and frequency rate parameters.
	\end{enumerate}
	We have used ``GA" package \cite{scrucca2013ga} from R software to implement GA.
	Initial population size for GA is fixed at 500, elite count at 20, fitness function is negative of first component error sum of squares.  The roulette
	wheel selection is used to determine parents for next generation of individuals. Probability of crossover is maintained at 0.9 and probability of mutation is set at 0.1. GA is then run for  100 generations.\\~\\
	After this step, we are left with only 1D optimization problem to find sequential plugin estimators of frequency parameters of the remaining components, which can be done easily. We have compared MSEs of the sequential plugin estimators using above implementation, with that of the PHAF estimators across 30 different SNR levels. MSEs are observed over 1000 replications. The sample size is \(N=500\) with the non-linear parameters same as that mentioned in \eqref{par_set}. GA algorithm in our case, took 56.6 seconds and grid search took 17.8 seconds to compute initial guess, on an average, over 30,000 replications.
	\begin{figure}
		\centering
		\subfloat[\textit{Estimators of $\alpha_1^0$. }
		]{\label{fig:alp4}{\includegraphics[width=0.82\linewidth]{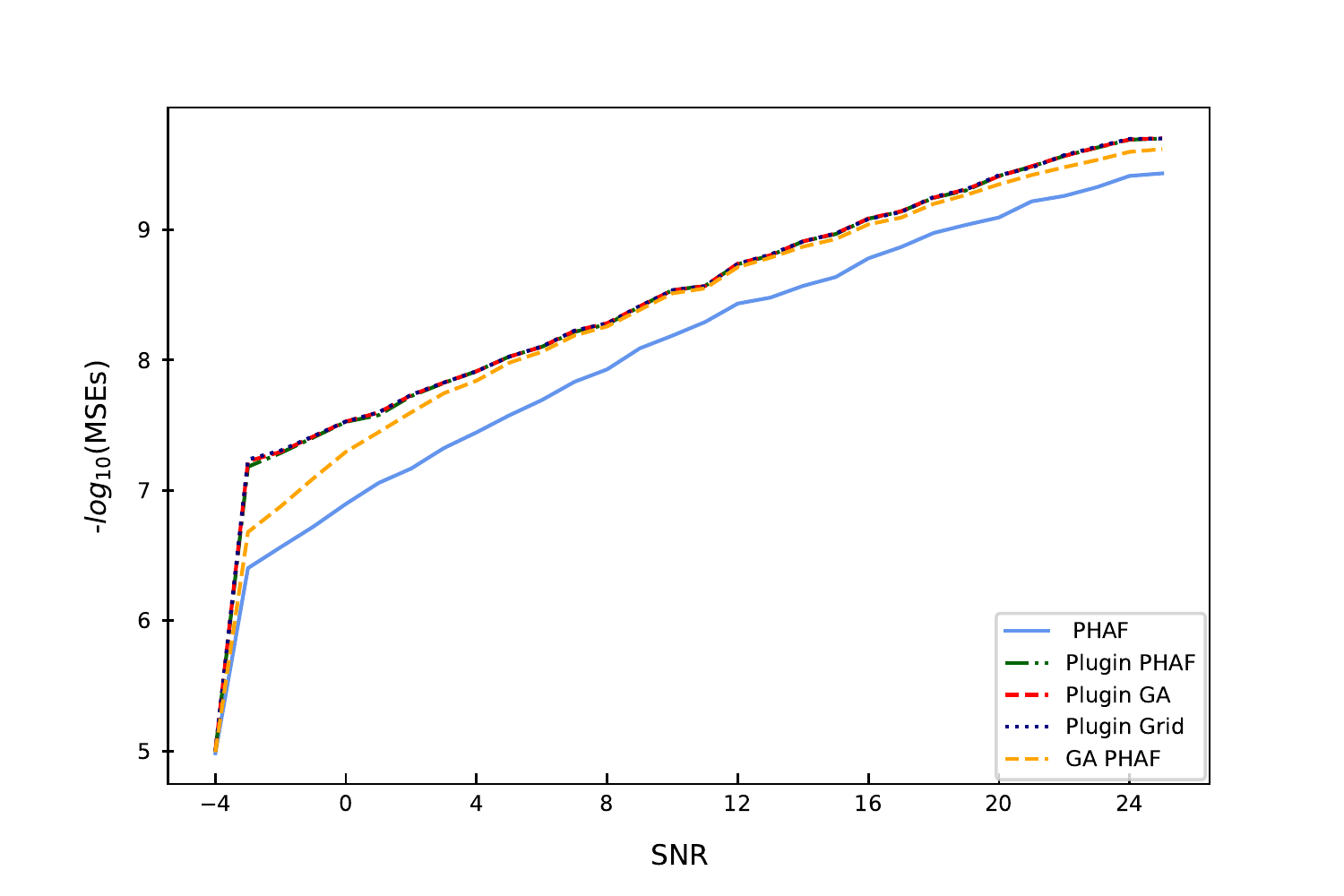}}}\qquad
		\subfloat[\textit{Estimators of $\alpha_2^0$.}]{\label{fig:beta4}{\includegraphics[width=0.82\linewidth]{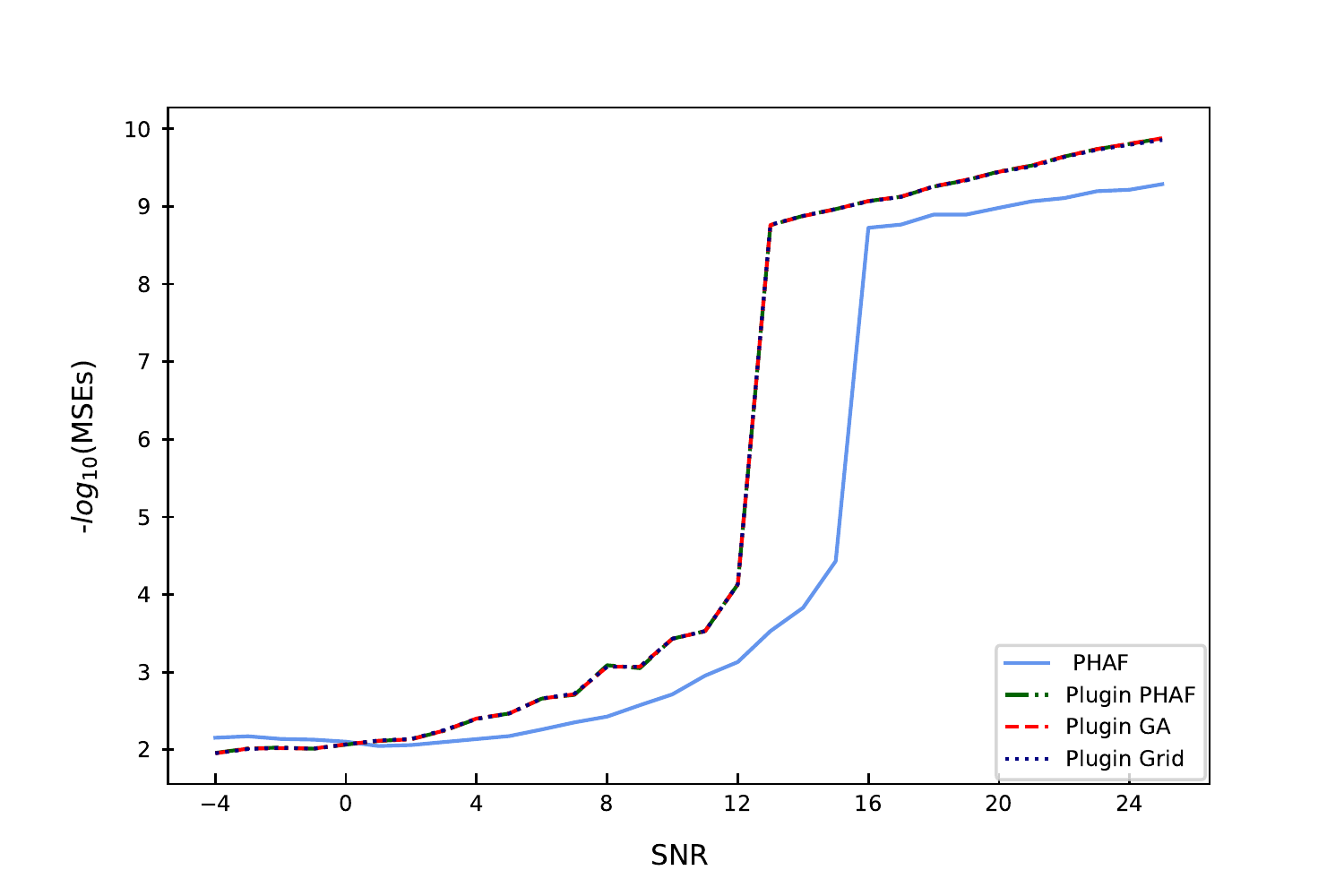}}}\\
		\caption{\textit{MSEs (on \(-\log_{10}\) scale) of PHAF estimators, sequential plugin estimators obtained by taking PHAF as initial guess (Plugin PHAF), sequential plugin estimators obtained by using GA as initial guess (Plugin GA), sequential plugin estimators obtained by grid search to get initial guess (Plugin Grid) and  GA with PHAF implementation (GA PHAF)  versus the SNR  for fixed sample size \(N=500\)}.}
		\label{plot_first_comp6}
	\end{figure}
	\begin{figure}
		\centering
		\subfloat[\textit{Estimators of $\alpha_3^0$. }
		]{\label{fig:alp5}{\includegraphics[width=0.82\linewidth]{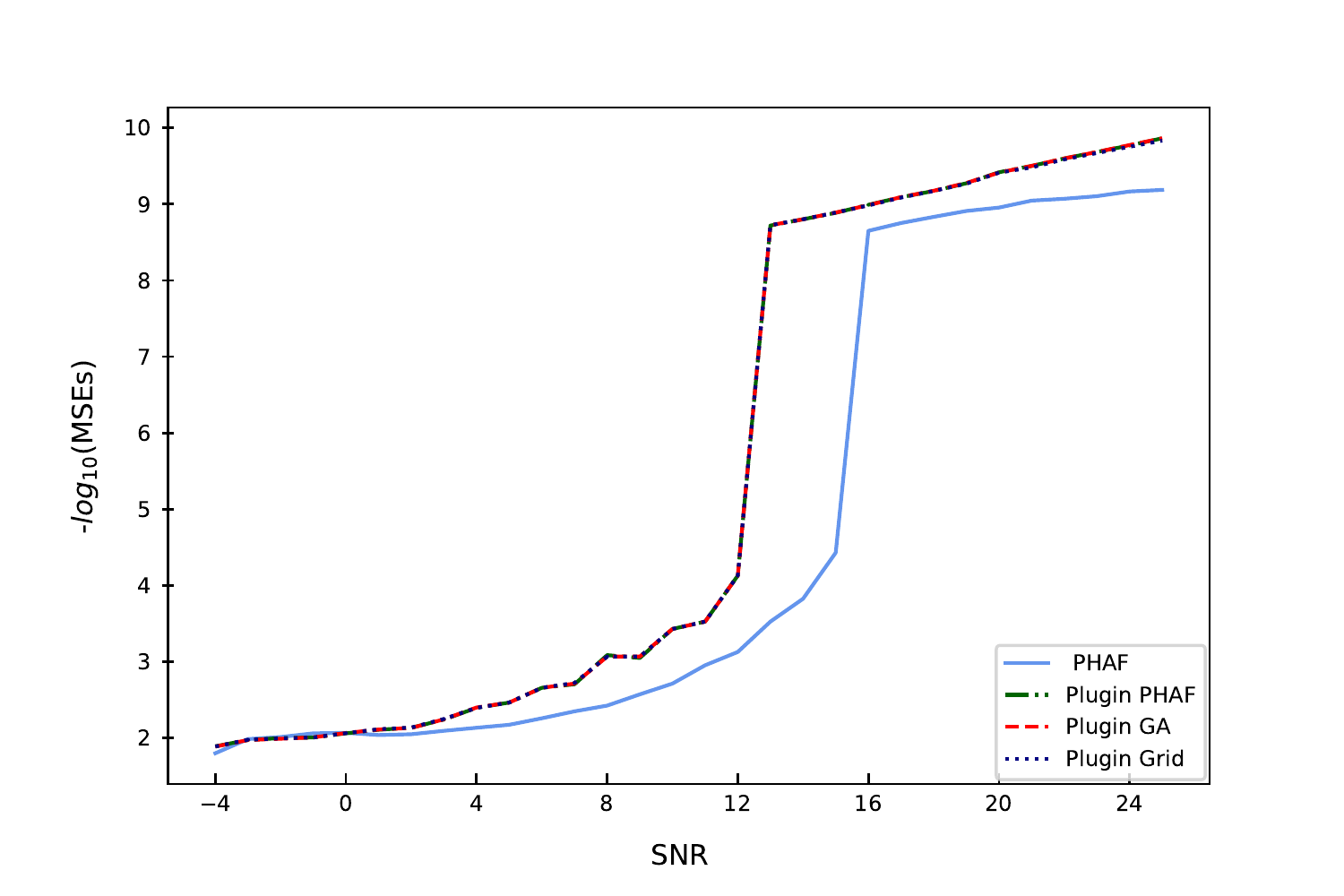}}}\qquad
		\subfloat[\textit{Estimators of $\alpha_4^0$.}]{\label{fig:beta5}{\includegraphics[width=0.82\linewidth]{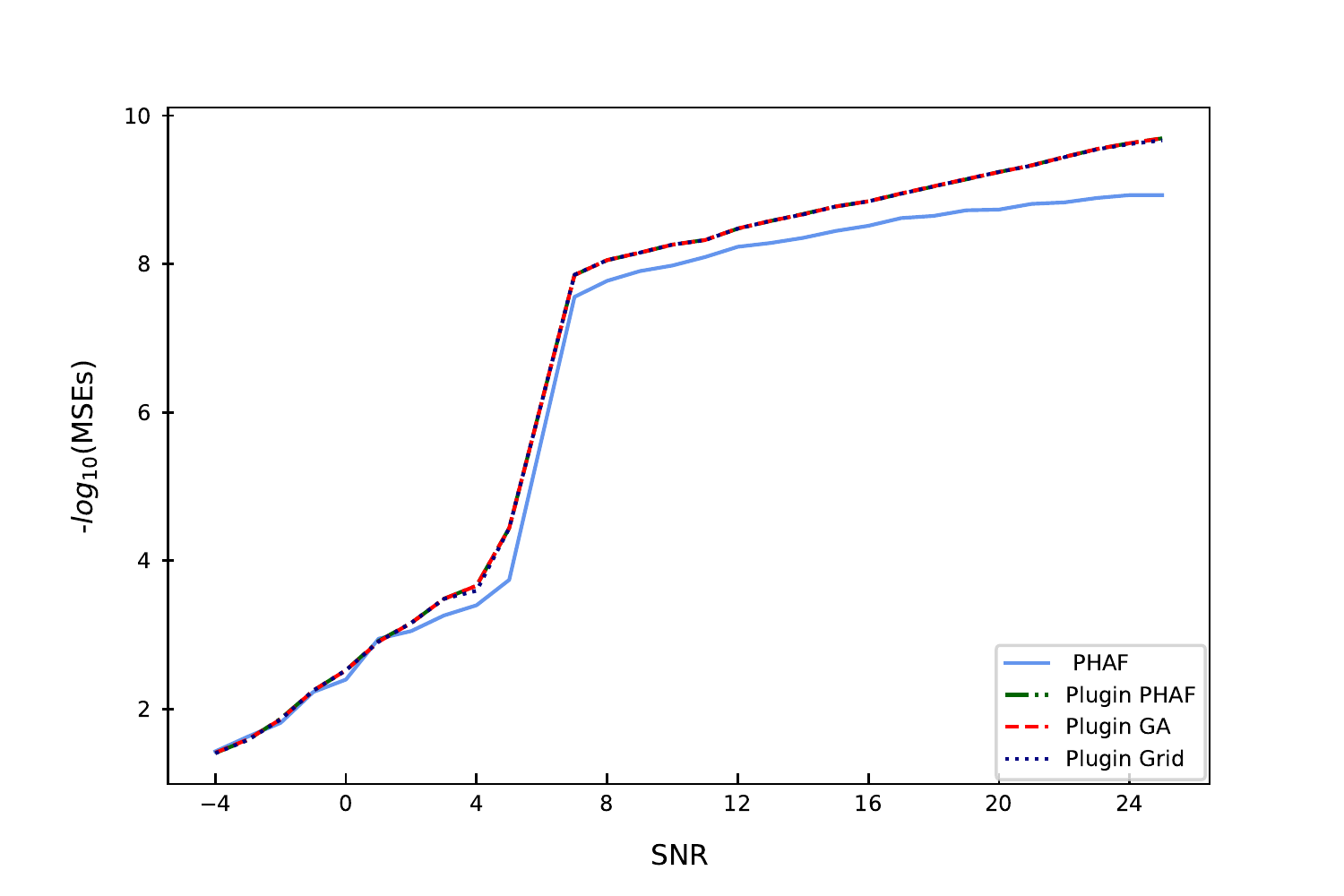}}}\\
		\caption{\textit{MSEs (on \(-\log_{10}\) scale) of PHAF estimators, sequential plugin estimators obtained by taking PHAF as initial guess (Plugin PHAF), sequential plugin estimators obtained by using GA as initial guess (Plugin GA) and sequential plugin estimators obtained by grid search to get initial guess (Plugin Grid)   versus the SNR  for fixed sample size \(N=500\)}.}
		\label{plot_first_comp7}
	\end{figure}
	\begin{figure}
		\centering
		\subfloat[\textit{Estimators of $\alpha_5^0$. }
		]{\label{fig:alp6}{\includegraphics[width=0.82\linewidth]{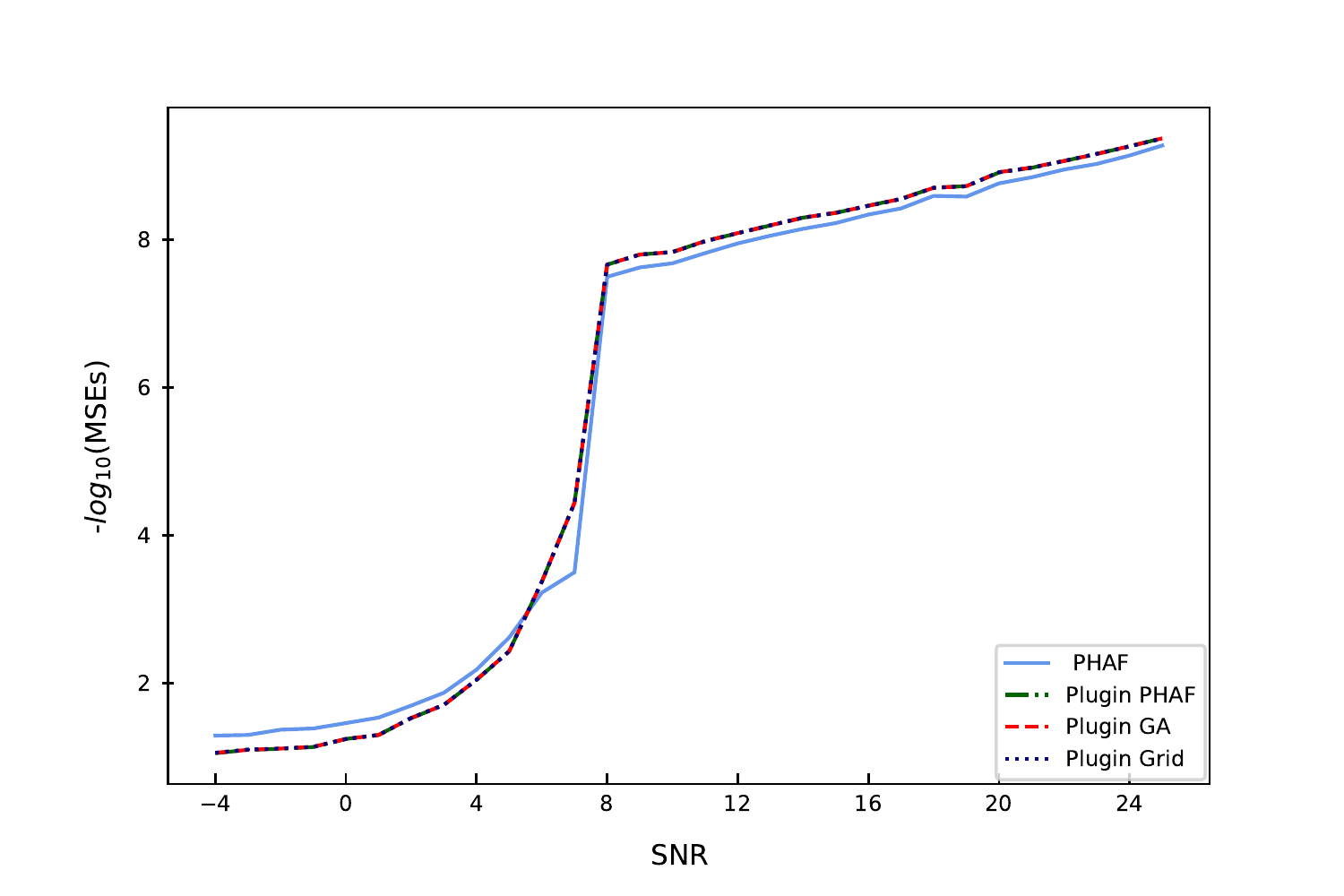}}}\qquad
		\subfloat[\textit{Estimators of $\beta^0$.}]{\label{fig:beta6}{\includegraphics[width=0.82\linewidth]{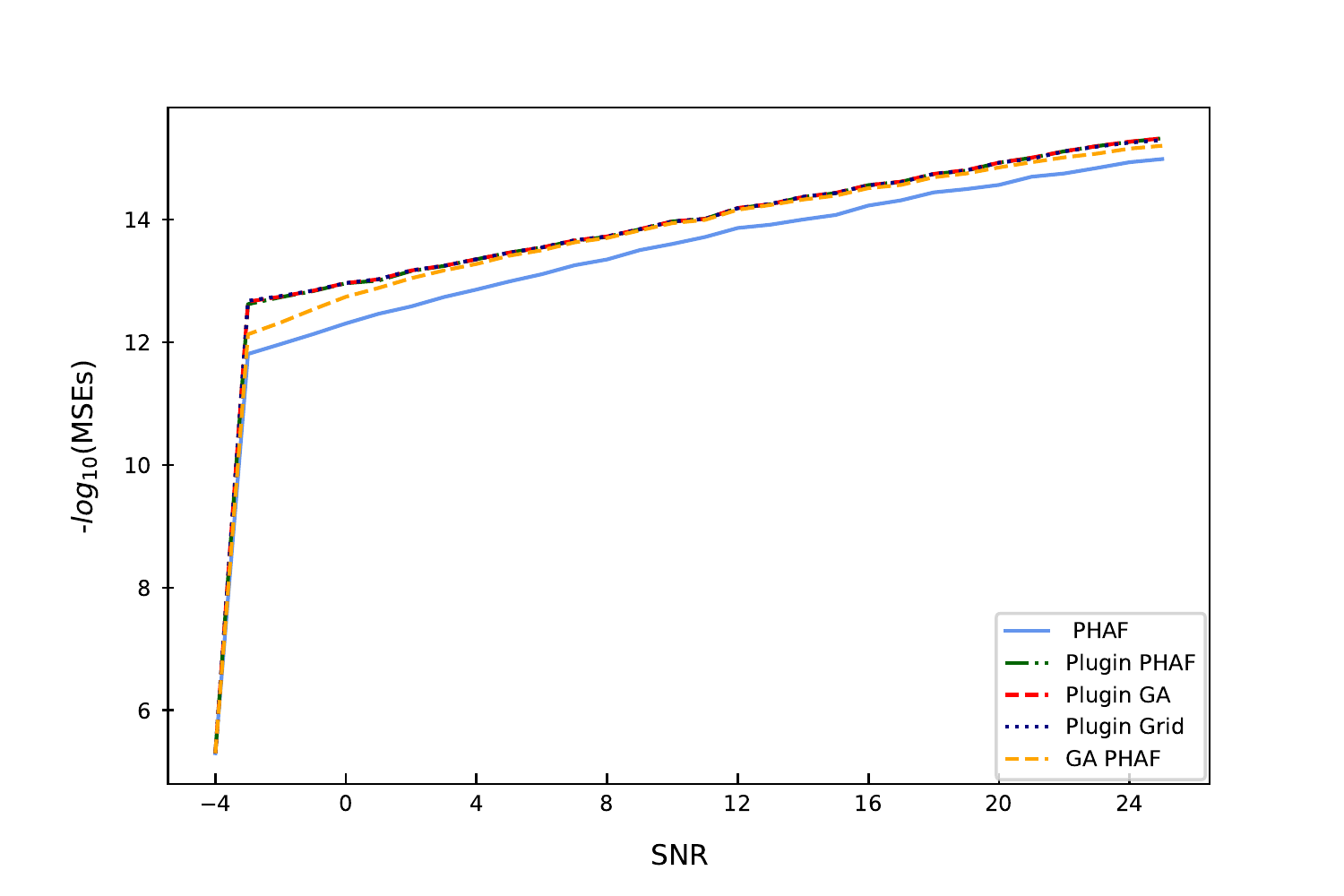}}}\\
		\caption{\textit{MSEs (on \(-\log_{10}\) scale) of PHAF estimators, sequential plugin estimators obtained by taking PHAF as initial guess (Plugin PHAF), sequential plugin estimators obtained by using GA as initial guess (Plugin GA), sequential plugin estimators obtained by grid search to get initial guess (Plugin Grid) and  GA with PHAF implementation (GA PHAF)  versus the SNR  for fixed sample size \(N=500\)}.}
		\label{plot_first_comp8}
	\end{figure}
	From Fig. \ref{plot_first_comp6}, \ref{plot_first_comp7} and \ref{plot_first_comp8}, it can be easily observed that:
	\begin{itemize}
		\item 	sequential plugin estimators of the first component parameters perform very well for even low SNR -3 dB;
		\item sequential plugin estimators for remaining component parameters attain better performance  around 12 dB;
		\item different implementation procedures to find initial guess   of the plugin estimators resulted in similar performance.
	\end{itemize} 
	We suggest to use PHAF directly as initial guess for the data analysis purpose, as it will be the least time consuming out of the proposed three methods and at the same time, it has comparable numerical performance as the remaining two options.
	{\subsection{Radar Data Analysis}\label{radar_data}
		We consider a radar data simulation obtained from \cite{ISAR_matlab_book} (Chapter 4) where,  110 point scatterers were placed to reproduce the outline of a fictitious airplane. Backscattered electric field
		for 64 different frequencies and 64 different aspects were collected. We further add i.i.d. normal errors with standard deviation 5 to the real and imaginary parts of the radar data, to make problem of estimation more realistic and difficult. The conventional small-bandwidth small-angle ISAR imaging algorithm is applied to obtain ISAR image of the airplane as shown in Fig. \ref{isar} with a certain level of contrast adjusted. Using motivation from \cite{Rao_2014}, we have fitted  twenty-component chirp with equal chirp rates model to the backscattered data observed in each range bin cell. After obtaining fitted data form these twenty-component chirp models in each range bins, we apply the conventional small-bandwidth small-angle ISAR imaging algorithm  to get the  ISAR image at the same contrast level as that of contaminated data Fig. \ref{isar} (a).\\~\\
		One can observe from Fig. \ref{isar} that using estimators with optimal rates of convergence can satisfactorily help in fitting the data obtained from backscattered electric field after hitting a target, whereas the ISAR image obtained using PHAF in Fig. \ref{isar} (d) is distorted and noisy. Combined estimator performs better than the plugin and PHAF based  estimators. Note that these plots are presented at the same level of contrast. Our proposed estimators can be combined further with a suitable contrasting algorithm, and then it can be used to improve ISAR imaging even in more difficult situations. We have proposed our estimators after assuming known number of chirp components, so 
		selecting twenty-component chirp model is subjective and existing methods like Bayesian information criteria (BIC) or maximum a posteriori probability (MAP) based approach fail to estimate appropriate number of components for such kind of radar datasets. We believe that the methods based on bayesian posterior inference methods will be effective in solving such problems. Hence, a comprehensive separate study  is required to study estimation of the number of components for such a dataset.  }	
	\begin{figure*}
		\centering
		\subfloat[{\textit{ISAR image of contaminated data.} }
		]{\label{fig:isar_contam}{\includegraphics[width=0.42\linewidth]{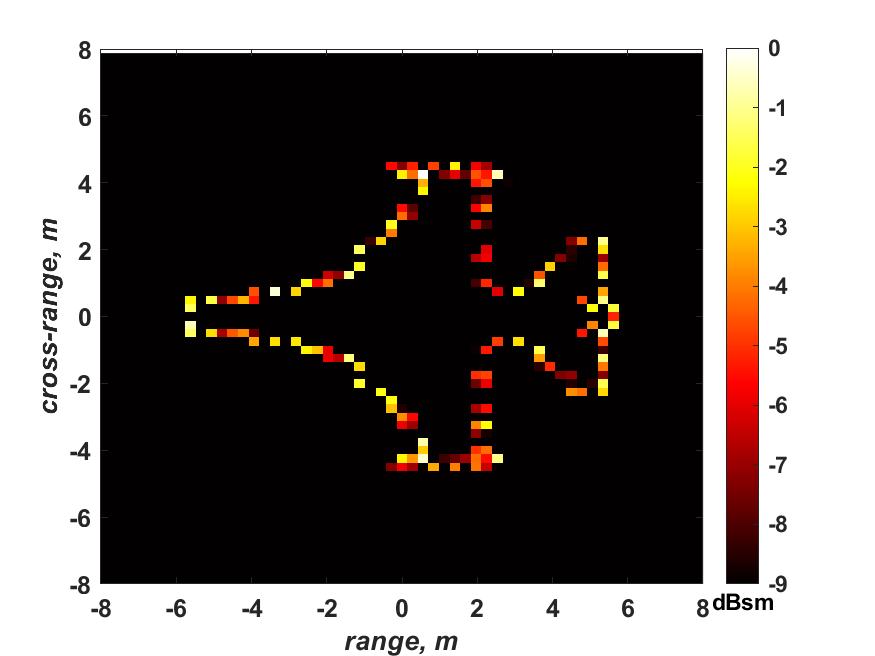}}}\qquad
		\subfloat[{\textit{ISAR image of data fitted usnig combined estimators.}}]{\label{fig:isar_combin}{\includegraphics[width=0.42\linewidth]{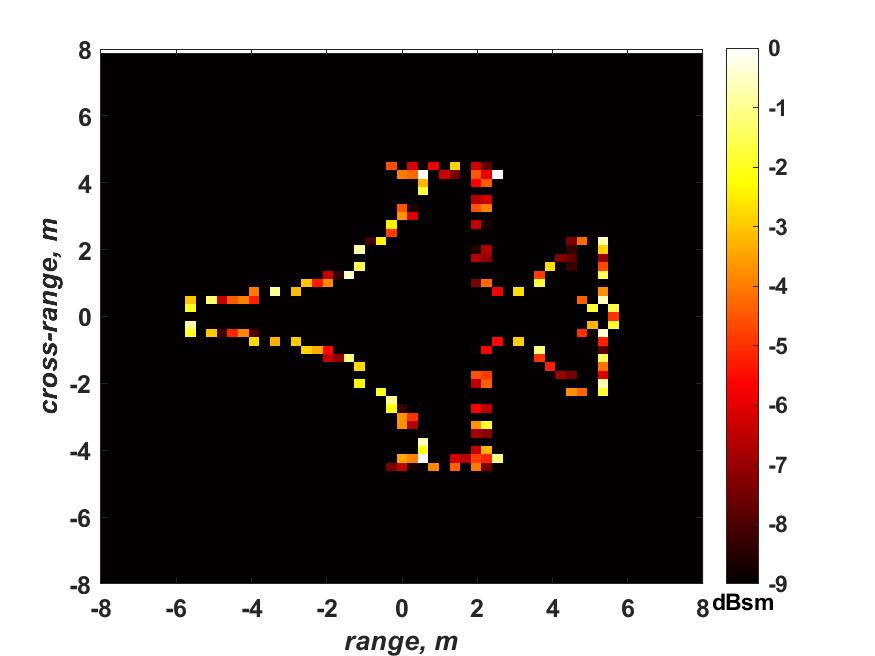}}}\\
		\subfloat[{\textit{ISAR image of data fitted usnig plugin estimators.} }
		]{\label{fig:isar_plugin}{\includegraphics[width=0.42\linewidth]{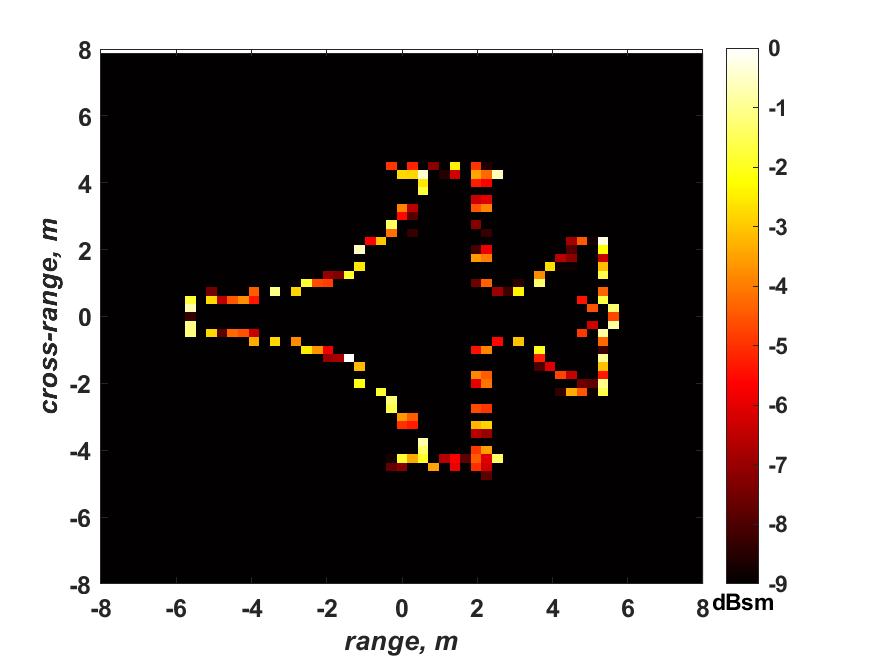}}}\qquad
		\subfloat[{\textit{ISAR image of data fitted usnig PHAF estimators.}}]{\label{fig:isar_phaf}{\includegraphics[width=0.42\linewidth]{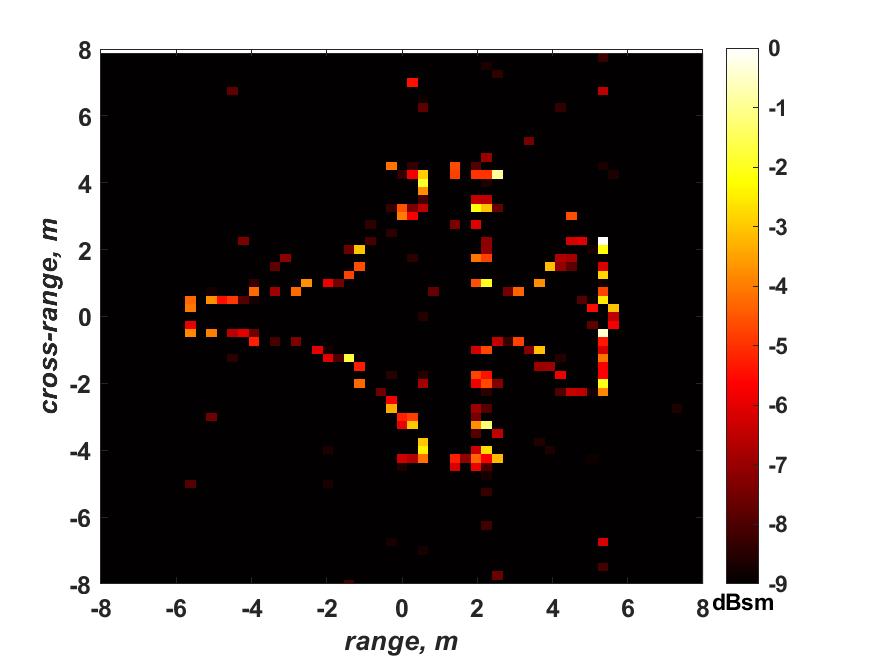}}}\\
		\caption{{\textit{ISAR image obtained from fitting twenty-component chirp model with equal chirp rates in each range bin signals.}}}
		\label{isar}
	\end{figure*}

	\section{Conclusion}\label{concl}
	We have established theoretical asymptotic properties of three estimators namely the LSEs, sequential combined estimators and sequential plugin estimators of the  parameters of a chirp signal model with equal chirp rates. We observe that among the three estimators, the sequential plugin estimator is computationally least burdensome and provides a computationally efficient alternative to the LSEs. Empirical results indicate stable performance of the sequential plugin estimator for the chirp rate parameter.  We further observe that sequential combined estimator of chirp rate parameter \(\beta^0\) is asymptotically optimal, with its asymptotic variance matching with that of the LSEs. {Analysis of radar signal data indicate that model \eqref{true model} can be used satisfactorily to analyse these type of signals and therefore, our proposed estimators have potential to be used further in radar applications. It is observed that fitting radar signal with model \eqref{true model} performs satisfactorily using our proposed estimators, specially combined estimator performs better than the plugin and PHAF based estimators. }Extensive numerical simulations support the derived theoretical asymptotic properties of the proposed estimators. {Future work may be directed towards estimating number of chirp signal components present in such radar signal data..}
	
	\section*{Acknowledgments}
	
	We are thankful to the Associate Editor Dr. Hasan Mir  for his encouragement, and the three unknown reviewers for their constructive suggestions. Their suggestions have helped improving the manuscript significantly.
	
	\section{Appendix A}\label{consis_proof}
	We  provide a very brief sketch of the proof of Theorem \ref{cons_tm} and \ref{asymp_norm_thm}.
	\subsection{Strong Consistency of LSEs}
	\begin{IEEEproof} 	Proof of \(\widehat{\bm{\theta}}\xrightarrow{a.s.}\bm{\theta}^0\) follows along the similar lines of proof of Theorem 1 of \cite{Kundu_2008} with appropriate changes.
	\end{IEEEproof} 
	\subsection{Strong Consistency of Sequential Combined Estimators}\label{cons_comb_prf}
	\begin{IEEEproof}
		Proof of \(\breve{\bm{\theta}}_k\xrightarrow{a.s.}\bm{\theta}_k^0\) follows along the similar lines of proof of Theorem 3, 4 and 5 of \cite{Lahiri_2015}. Note that \(\breve{\beta}=\cfrac{1}{\displaystyle \sum_{k=1}^{p}(\breve{A}_k^{0^2}+\breve{B}_k^{0^2})}\displaystyle \sum_{k=1}^{p}(\breve{A}_k^{0^2}+\breve{B}_k^{0^2})\breve{\beta}_k\) is a continuous function of strongly consistent estimators and hence, by continuous mapping theorem, \(\breve{\beta}\) is strongly consistent for \(\beta^0\).
	\end{IEEEproof}
	\subsection{Strong Consistency of Sequential Plugin Estimators}
	\begin{IEEEproof}	We know that first component sequential plugin estimators are same as first component sequential combined estimators, so \[\widetilde{\bm{\theta}}_1\xrightarrow{a.s.}(A_1^0,B_1^0,\alpha_1^0,\beta^0)^\top\] follows from previous argument (see Section \ref{cons_comb_prf}).
		We  provide a sketch of proof of consistency for \(\widetilde{\bm{\theta}}_2\), and strong consistency of \(\widetilde{\bm{\theta}}_k\) will follow similar to it for \(k\geq3\). First consider the following lemma:
		\begin{lemma}\label{lemm_1}
			Consider the set \(S_{c}=\{\bar{\bm{\theta}}_2:\bar{\bm{\theta}}_2\in \bar{\bm{\Theta}}_2, |\bar{\bm{\theta}}_2-\bm{\theta}_2^0|\geq c\} \), where \(\bar{\bm{\theta}}_2=(A_2,B_2,\alpha_2)^\top\),  \(\bm{\theta}_2^0=(A_2^0,B_2^0,\alpha_2^0)^\top\) and \(\bar{\bm{\Theta}}_2=[-M,M]\times[-M,M]\times(0,2\pi)\). If \\
			\(\liminf \inf\limits_{\bar{\bm{\theta}}_2\in S_c}\cfrac{1}{N}\Big(Q_2(\bar{\bm{\theta}}_2)-Q_2({\bm{\theta}}_2^0)\Big)>0\hspace{5pt} a.s.\) for any \(c>0\), then \(\widetilde{\bm{\theta}}_2\), the minimizer of \(Q_2(\bar{\bm{\theta}}_2)\), is a strongly consistent estimator of  \(\bm{\theta}_2^0\). 			
		\end{lemma}\begin{proof}
			It follows along the same lines of proof of Lemma 1 of \cite{Wu_1981}.
		\end{proof}
		Now, we need to show that \(\liminf \inf\limits_{\bar{\bm{\theta}}_2\in S_c}\cfrac{1}{N}\Big(Q_2(\bar{\bm{\theta}}_2)-Q_2({\bm{\theta}}_2^0)\Big)>0\hspace{5pt} a.s.\). We will use the result from  \cite{Lahiri_2015} for first component sequential plugin estimator, i.e.,\\ \(\widetilde{\beta}=\beta^0+o(1/N^2)\). Expanding the difference 
		\(\Big(Q_2(\bar{\bm{\theta}}_2)-Q_2({\bm{\theta}}_2^0)\Big)\), and following the arguments of proof of Theorem 3 in \cite{Lahiri_2015}, it follows that \(\liminf \inf\limits_{\bar{\bm{\theta}}_2\in S_c}\cfrac{1}{N}\Big(Q_2(\bar{\bm{\theta}}_2)-Q_2({\bm{\theta}}_2^0)\Big)>0\hspace{5pt} a.s.\) holds for any \(c>0\).
	\end{IEEEproof}
	\section{Appendix B}\label{normal_proof}
	\subsection{Asymptotic Normality of LSEs}\label{lses_asymp_norm_proof}
	\begin{IEEEproof} 
		We will follow the notations from subsection \ref{LSEs_methd}. Using multivariate Taylor Series expansion, we have:
		\[\bm{Q}^{\prime}(\widehat{\bm{\theta}})-\bm{Q}^{\prime}(\bm{\theta}^0)=\bm{Q}^{\prime\prime}(\hat{\hat{\bm{\theta}}})(\widehat{\bm{\theta}}-\bm{\theta}^0), \mbox{}\]
		where, 	\(\hat{\hat{\bm{\theta}}}\) is a point between \(\hat{\bm{\theta}}\) and \(\bm{\theta}^0\).
		\[\text{Above yields } \bm{D}^{-1}(\widehat{\bm{\theta}}-\bm{\theta}^0)=-[\bm{DQ}^{\prime\prime}(\hat{\hat{\bm{\theta}}})\bm{D}]^{-1} \bm{DQ}^{\prime}(\bm{\theta}^0), \]
		\( \mbox{ where } \bm{Q}^{\prime}(\widehat{\bm{\theta}})=\bm{0}\) and \(\bm{D}\) matrix is defined in Theorem \ref{asymp_norm_thm}.
		Let us now consider the derivative vector  \(	\bm{DQ}^{\prime}(\bm{\theta}^0)\). By using the central limit theorem of stochastic processes (see \cite{Fuller_2009}), we obtain asymptotic distribution of \(\bm{DQ}^{\prime}(\bm{\theta}^0)\) as multivariate normal with mean vector \(\bm{0}\) and variance co-variance matrix  $2c\sigma^2\bm{\Sigma}_1^{-1}$ which can be represented in block matrix form as:
		\(2c\sigma^2\begin{bmatrix}
			\bm{\Sigma}_{11}&	\bm{\Sigma}_{12}\\
			\bm{\Sigma}_{21}&	\bm{\Sigma}_{22}
		\end{bmatrix}.\) Inversion of this block-matrix will provide elements of \(\bm{\Sigma}_1\). 
		Here, \(\bm{\Sigma}_{11}\) is the identity matrix of order \(2p\times 2p\), 	\(
		\bm{\Sigma}_{21}=	\bm{\Sigma}_{12}^\top\), \begin{align*}	\begin{small} 
				\bm{\Sigma}_{12}=\begin{bmatrix}
					B_1^0/2&  0&0&0&\cdots&0&                         B_1^0/3\\   
					- A_1^0/2&  0&0&0&\cdots&0&                         -A_1^0/3\\   
					0&		 B_2^0/2&0& 0&\cdots&0&                         B_2^0/3\\   
					0&	- A_2^0/2& 0&0&\cdots&0&                         -A_2^0/3\\
					\vdots&\vdots&\vdots&\vdots&\vdots&\vdots&\vdots\\
					0&0&		 0&0&		\cdots&B_p^0/2&                        B_p^0/3\\
					0&0&		 0&0&		\cdots&-A_p^0/2&                        -A_p^0/3\\
				\end{bmatrix}_{\big(2p\times(p+1)\big)},
			\end{small} 
		\end{align*}
		and for \(k=1,2,\ldots,p\), the	\(k^{th}\) row of  \(\bm{\Sigma}_{22}\)   contains \(\cfrac{\Big(A_k^{0^2} + B_k^{0^2}\Big)}{3}\) at \(k^{th}\) column and \(\cfrac{\Big(A_k^{0^2} + B_k^{0^2}\Big)}{4}\) at \((p+1)^{th}\) column. The \((p+1)\times(p+1)^{th}\) element of  \(\bm{\Sigma}_{22}\)  is \(\displaystyle \sum_{k=1}^{p}\cfrac{\Big(A_k^{0^2} + B_k^{0^2}\Big)}{5}\). Remaining elements of \((p+1)^{th}\) row are obtained by symmetry. The rest of the elements of \(\bm{\Sigma}_{22}\)  are 0. Using expression of \(\bm{\Sigma}_1^{-1}\), we can easily calculate each term of the matrix  \(\bm{\Sigma}_1\).\\
		It can be seen that $[\bm{DQ}^{\prime\prime}(\hat{\hat{\bm{\theta}}})\bm{D}]^{-1}\xrightarrow{n\rightarrow \infty}\bm{\Sigma}_1$, where $2c\sigma^2\bm{\Sigma}_1^{-1}$ is the variance co-variance matrix of $	\bm{DQ}^{\prime}(\bm{\theta}^0)$. By simple calculations, we get following results:
		\begin{align}\label{asymp_lses_bet_alph}
			AsymVar\Big(N^{-5/2}(\widehat{\beta}-\beta^0)\Big)&=\cfrac{360c\sigma^2}{\displaystyle \sum_{k=1}^{p}\Big(A_k^{0^2} + B_k^{0^2}\Big)},
		\end{align} 
		\begin{align} \label{asymp_lses_bet_alph2}
			AsymVar\Big(N^{-3/2}(\widehat{\alpha}_k-\alpha_k^0)\Big)&=\cfrac{360c\sigma^2}{\displaystyle \sum_{k=1}^{p}\Big(A_k^{0^2}+ B_k^{0^2}\Big)}\\&+\cfrac{24c\sigma^2}{\Big(A_k^{0^2} + B_k^{0^2}\Big)}.	\nonumber
		\end{align}
	\end{IEEEproof}
	\subsection{ Asymptotic Normality of Sequential Combined Estimators:}\label{seq_comb_asym_norm_proof}
	\begin{IEEEproof} 
		We will follow the notations from subsection \ref{seq_comb_methd}. 	If we define $\bm{\theta}_1=({A}_1,{B}_1,\alpha_1,\beta_1)^\top$, then for the following sum of squares,
		\begin{align*} Q_1(\bm{\theta}_1)&=\displaystyle \sum_{n=1}^{N}\big(y(n) -A_1\cos(\alpha_1 n+\beta_1 n^2)\\&-B_1\sin(\alpha_1 n+\beta_1 n^2)\big)^2.
		\end{align*}
		Using multivariate Taylor Series expansion, we have:
		\[\bm{Q}_1^{\prime}(\breve{\bm{\theta}}_1)-\bm{Q}_1^{\prime}(\bm{\theta}_1^0)=\bm{Q}_1^{\prime\prime}(\breve{\breve{\bm{\theta}}}_1)(\breve{\bm{\theta}}_1-\bm{\theta}_1^0), \mbox{}\]
		where, 	\(\breve{\breve{\bm{\theta}}}_1\) is a point between \(\breve{\bm{\theta}}_1\) and \(\bm{\theta}_1^0\). For \(\bm{D}_1^{-1}=diag(N^{1/2},N^{1/2},N^{3/2},N^{5/2})\), this yields,
		\[ \bm{D}_1^{-1}(\breve{\bm{\theta}}_1-\bm{\theta}_1^0)=-[\bm{D}_1\bm{Q}_1^{\prime\prime}(\breve{\breve{\bm{\theta}}}_1)\bm{D}_1]^{-1} \bm{D}_1\bm{Q}_1^{\prime}(\bm{\theta}_1^0), \]
		where,  \(\bm{Q}_1^{\prime}(\breve{\bm{\theta}}_1)=\bm{0}.\)
		We first calculate the derivative vector \(\bm{Q}_1^{\prime}(\bm{\theta}_1^0)\) as:
		\begin{align*}
			\cfrac{\partial Q_1(\bm{\theta}_1)}{\partial{A_1}}\bigg|_{\bm{\theta}_1=\bm{\theta}_1^0}&=
			-2\displaystyle \sum_{n=1}^{N}\big(y(n) -A_1^0\cos(\alpha_1^0 n+\beta^0 n^2)\\&-B_1^0\sin(\alpha_1^0 n+\beta^0 n^2)\big)\cos(\alpha_1^0 n+\beta^0 n^2)\\
			&=
			-2\displaystyle \sum_{n=1}^{N}\big(X(n) +A_2^0\cos(\alpha_2^0 n+\beta^0 n^2)\\&+B_2^0\sin(\alpha_2^0 n+\beta^0 n^2)\big)\cos(\alpha_1^0 n+\beta^0 n^2)\\&= 
			-2\displaystyle \sum_{n=1}^{N}\bigg(X(n)\cos(\alpha_1^0 n+\beta^0 n^2)\bigg)+o(\sqrt{N}).
		\end{align*}
		Last equality above, is observed using conjectures mentioned in \cite{Lahiri_2015} and Lemma 2 of \cite{Grover_2018}. 
		Similar calculations hold for other  derivative terms. It can then be shown that the first component estimators follow asymptotic properties same as   the LSEs of  mono-component chirp model parameters. If we consider second component estimators  $\breve{\bm{\theta}}_2=(\breve{A}_2,\breve{B}_2,\breve{\alpha}_2,\breve{\beta}_2)^\top$, and  define
		\begin{small}
			\[\breve{y_1}(n)=	y(n)-\breve{A}_1\cos(\breve{\alpha}_1 n+\breve{\beta}_1 n^2)-\breve{B}_1\sin(\breve{\alpha}_1 n+\breve{\beta}_1 n^2), \mbox{ and  }\]
			\[Q_2(\bm{\theta}_2)=\displaystyle \sum_{n=1}^{N}\big(\breve{y_1}(n) -A_2\cos(\alpha_2 n+{\beta}_2 n^2)-B_2\sin(\alpha_2 n+{\beta}_2 n^2)\big)^2.\]\end{small}
		Derivative vector of  \(Q_2(\bm{\theta}_2)\) at \({\bm{\theta}_2=\bm{\theta}_2^0}\) will have following elements:
		\begin{align*}
			\cfrac{\partial Q_2(\bm{\theta}_2)}{\partial{A_2}}\bigg|_{\bm{\theta}_2=\bm{\theta}_2^0}=&
			-2\displaystyle \sum_{n=1}^{N}\bigg(X(n)\cos(\alpha_2^0 n+\beta^0 n^2)\bigg) +o(\sqrt{N}),\\
			\cfrac{\partial Q_2(\bm{\theta}_2)}{\partial{\alpha_2}}\bigg|_{\bm{\theta}_2=\bm{\theta}_2^0}=&
			-2\displaystyle \sum_{n=1}^{N}\bigg(X(n)\big(A_2^0n\sin(\alpha_2^0 n+{\beta}^0 n^2)\\&-B_2^0n\cos(\alpha_2^0 n+{\beta}^0 n^2)\big)\bigg) +o(N\sqrt{N}),	
		\end{align*} 
		\begin{align*} 
			\cfrac{\partial	Q_2(\bm{\theta}_2)}{\partial{\beta}}\bigg|_{\bm{\theta}_2=\bm{\theta}_2^0}&=
			-2\displaystyle \sum_{n=1}^{N}\bigg(X(n)\big(A_2^0n^2\sin(\alpha_2^0 n+{\beta}^0 n^2)\\&-B_2^0n^2\cos(\alpha_2^0 n+{\beta}^0 n^2)\big)\bigg) +o(N\sqrt{N}).
		\end{align*}
		Arguing along the same lines of proof of the first component, we get for \(k\geq 1\),  $\breve{\bm{\theta}}_k$ asymptotically follows Normal distribution with variance covariance matrix  $$\cfrac{2c\sigma^2}{A_k^{0^2}+B_k^{0^2}}\begin{small} \begin{bmatrix}
				A_k^{0^2}+9B_k^{0^2}&-8A_k^0B_k^0&-36B_k^0&30B_k^0\\
				-8A_k^0B_k^0&9A_k^{0^2}+B_k^{0^2}&36A_k^0&-30A_k^0	\\
				-36B_k^0&36A_k^0&192&-180	\\
				30B_k^0&-30A_k^0&-180&180	
			\end{bmatrix}.\end{small} $$
		Therefore, asymptotic variance of sequential combined estimators of \(\alpha_k^0\)  is given by:
		\begin{equation}
			AsymVar\Big(N^{-3/2}(\breve{\alpha}_k-\alpha_k^0)\Big)=\cfrac{384c\sigma^2}{\Big(A_k^{0^2} + B_k^{0^2}\Big)}.
		\end{equation}
		Note that different component sequential combined  estimators are asymptotically independent.
		So, the final  estimator of $\beta^0$ is given by minimizing the variance of linear combination:
		$$\breve{\beta}=\displaystyle \sum_{k=1}^{p}l_k\breve{\beta}_k,\hspace{20pt}l_k\geq0,\displaystyle \sum_{k=1}^{p}l_k=1$$
		Solving above yields $l_k=\cfrac{(A_k^{0^2}+B_k^{0^2})}{\displaystyle \sum_{k=1}^{p}(A_k^{0^2}+B_k^{0^2})}$.\\~\\We can use consistent estimators of $l_k$ as they are functions of amplitude parameters. After simple calculations, it can be shown that asymptotic variance of $\breve{\beta}$ matches with that of LSEs. 
	\end{IEEEproof} %
	\subsection{ Asymptotic Normality of Sequential Plugin Estimators} 	
	\begin{IEEEproof} 
		Note that sequential plugin estimators of  first component parameters	are same as the sequential combined estimators. Thus, we have	\(\bm{D}_1^{-1}\Big(\widetilde{\bm{\theta}}_1-\bar{\bm{\theta}}_1^0\Big)\) converging in distribution to normal  with asymptotic mean \(\bm{0}\) and variance covariance matrix
		
		$$\cfrac{2c\sigma^2}{A_1^{0^2}+B_1^{0^2}}\begin{small}\begin{bmatrix}
				A_1^{0^2}+9B_1^{0^2}&-8A_1^0B_1^0&-36B_1^0&30B_1^0\\
				-8A_1^0B_1^0&9A_1^{0^2}+B_1^{0^2}&36A_1^0&-30A_1^0	\\
				-36B_1^0&36A_1^0&192&-180	\\
				30B_1^0&-30A_1^0&-180&180	
			\end{bmatrix}.\end{small} $$
		where \(\bm{D}_1^{-1}=diag(N^{1/2},N^{1/2},N^{3/2},N^{5/2})\). 
		Now for the second component estimators,  $\widetilde{\bm{\theta}}_2=(\widetilde{A}_2,\widetilde{B}_2,\widetilde{\alpha}_2)^\top$ of \(\bm{\theta}_2^0=(A_2^0,B_2^0,\alpha_2^0)^\top\),  we update the data as:\\
		\(\widetilde{y_1}(n)=	y(n)-\widetilde{A}_1\cos(\widetilde{\alpha}_1 n+\widetilde{\beta} n^2)-\widetilde{B}_1\sin(\widetilde{\alpha}_1 n+\widetilde{\beta} n^2)\)
		and define the following sum of squares:\begin{align*} Q_2(\bar{\bm{\theta}}_2)=\displaystyle \sum_{n=1}^{N}\big(\widetilde{y_1}(n) &-A_2\cos(\alpha_2 n+\widetilde{\beta} n^2)\\&-B_2\sin(\alpha_2 n+\widetilde{\beta} n^2)\big)^2.
		\end{align*} 
		Now, we calculate the elements of derivative vector \(\bm{Q}^\prime_2(\bar{\bm{\theta}}_2^0)\)  in following steps:
		\begin{align*}
			\cfrac{\partial Q_2(\bar{\bm{\theta}}_2)}{\partial{A_2}}\bigg|_{\bar{\bm{\theta}}_2=\bar{\bm{\theta}}_2^0}=&
			-2\displaystyle \sum_{n=1}^{N}\bigg(X(n)\cos(\alpha_2^0 n+{\beta^0} n^2)\bigg)\\&+B_2^0N^3(\widetilde{\beta}-\beta^0)/3+o(\sqrt{N}).
		\end{align*}
		Similarly, it can be shown that \begin{align*}
			\cfrac{\partial Q_2(\bar{\bm{\theta}}_2)}{\partial{B_2}}\bigg|_{\bar{\bm{\theta}}_2=\bar{\bm{\theta}}_2^0}=&
			-2\displaystyle \sum_{n=1}^{N}\bigg(X(n)\sin(\alpha_2^0 n+{\beta^0} n^2)\bigg)\\&-A_2^0N^3(\widetilde{\beta}-\beta^0)/3+o(\sqrt{N}),	
		\end{align*} 
		\begin{align*}
			&	\cfrac{\partial Q_2(\bar{\bm{\theta}}_2)}{\partial{\alpha_2}}\bigg|_{\bar{\bm{\theta}}_2=\bar{\bm{\theta}}_2^0}=
			2\displaystyle \sum_{n=1}^{N}\bigg(nX(n)\Big(A_2^0\sin(\alpha_2^0 n+{\beta^0} n^2)\\&-B_2^0\cos(\alpha_2^0 n+{\beta^0} n^2)\Big)\bigg)\\&+\Big(A_2^{0^2}+B_2^{0^2}\Big)N^4(\widetilde{\beta}-\beta^0)/4+o(N\sqrt{N}).	
		\end{align*} 
		$\bm{D}_2\bm{Q}_2^{\prime\prime}(\bar{\bm{\theta}}_2^0)\bm{D}_2 \xrightarrow{n \rightarrow \infty}  \begin{small}\begin{bmatrix}
				1&0&B_2^0/2\\
				0&1&-A_2^0/2\\
				B_2^0/2&-A_2^0/2&(A_2^{0^2}+B_2^{0^2})/3
		\end{bmatrix}\end{small} =\bar{\bm{\Sigma}}_2$, where \(\bm{D}_2^{-1}=diag(N^{1/2},N^{1/2},N^{3/2})\), and 
		$\Big[\bm{D}_2\bm{Q}_2^{\prime\prime}(\bar{\bm{\theta}}_2^0)\bm{D}_2\Big]^{-1} \xrightarrow{n \rightarrow \infty} $ $$ \cfrac{1}{A_2^{0^2}+B_2^{0^2}}\begin{small} \begin{bmatrix}
				A_2^{0^2}+4B_{2}^{0^2}&-3A_2^0B_2^0&-6B_2^0\\
				-3A_2^0B_2^0&4A_2^{0^2}+B_2^{0^2}&6A_2^0\\
				-6B_2^0&6A_2^0&12
			\end{bmatrix}.
		\end{small} $$
		Derivative vectors properties  for any $k^{th}$ component with $k=3,4,\ldots,p$  can be obtained similarly. Some important observations on asymptotic variances and covariances are provided in the supplementary material (page 1, 2 and 3). 
		Collecting these variances and covariances, we write  $2c\sigma^2\bm{\Sigma}_3^k$ as the asymptotic variance covariance matrix  of $\bm{D}_2^{-1}\bm{Q}_k^{\prime}(\bar{\bm{\theta}}_k^0)$, then asymptotic variance covariance matrix  of $\widetilde{\bm{\theta}}_k$ is given by:
		$$2c\sigma^2\bar{\bm{\Sigma}}_k^{-1} \bm{\Sigma}_3^k\bar{\bm{\Sigma}}_k^{-1},$$
		where \(\lim_{n \rightarrow \infty}	\bm{D}_2\bm{Q}_k^{\prime\prime}(\bar{\bm{\theta}}_k^0)\bm{D}_2 =\bar{\bm{\Sigma}}_k \). 
		After some straight-forward calculations, we observe that 
		\begin{align} 
			AsymVar(\widetilde{\alpha}_k)&=\cfrac{24c\sigma^2}{\big (A_k^{0^2}+B_k^{0^2}\big)}+\cfrac{360c\sigma^2}{\big (A_1^{0^2}+B_1^{0^2}\big)},\\
			AsymVar(\widetilde{\beta})&=\cfrac{360c\sigma^2}{\big (A_1^{0^2}+B_1^{0^2}\big)},
		\end{align} 
		\begin{align} 
			AsymVar(\widetilde{A}_k)\nonumber&=\cfrac{2c\sigma^2}{\big (A_k^{0^2}+B_k^{0^2}\big)}\bigg[\big (A_k^{0^2}+4B_k^{0^2}\big)\\&+\cfrac{5B_k^{0^2}\big (A_k^{0^2}+B_k^{0^2}\big)}{\big (A_1^{0^2}+B_1^{0^2}\big)}\bigg], 
		\end{align} 
		\begin{align} \label{last_eqn}
			AsymVar(\widetilde{B}_k)\nonumber&=\cfrac{2c\sigma^2}{\big (A_k^{0^2}+B_k^{0^2}\big)}\bigg[\big (4A_k^{0^2}+B_k^{0^2}\big)\\&+\cfrac{5A_k^{0^2}\big (A_k^{0^2}+B_k^{0^2}\big)}{\big (A_1^{0^2}+B_1^{0^2}\big)}\bigg] .\end{align} 
		Asymptotic covariance matrix $2c\sigma^2\bm{\Sigma}_{1k}$ between $\widetilde{\bm{\theta}}_1$ and $\widetilde{\bm{\theta}}_k$ is given by
		
		\[\cfrac{2c\sigma^2}{\big(A_1^{0^2}+B_1^{0^2}\big)}\begin{bmatrix}
			5B_1^0B_k^0&-5B_1^0A_k^0&-30B_1^0\\
			-5A_1^0B_k^0&5A_1^0A_k^0&30A_1^0\\
			-30B_k^0&30A_k^0&180\\
			30B_k^0&-30A_k^0&-180
		\end{bmatrix}.\]
		
		Asymptotic covariance between $k^{th}$ and $j^{th}$ components estimators for $1<k< j\leq p,$ can be obtained by using  results provided in the supplementary material (page 3).
		
	\end{IEEEproof} 
	A short comparison between asymptotic variance of sequential plugin and sequential combined estimators of   $\alpha_k^0$, $A_k^0$ and $B_k^0$, for $k\geq2$ has also been put in the supplementary material (page 3) for the sake of completeness of the study.
	\ifCLASSOPTIONcaptionsoff
	\newpage
	\fi
	
	\text{}\vskip -2. cm\text{}
	\begin{IEEEbiographynophoto}{Abhinek Shukla	}
		completed B.Sc Hons. degree in Statistics and M.sc degree in Statistics  from Banaras Hindu University in the period of 2013-2018. He joined IIT Kanpur, India, as a research scholar under the supervision of Prof. Amit Mitra and Prof. Debasis Kundu. His current research interests are parameter estimation and detection of polynomial phase signal models. He has also worked in improving inference based on averaged stochastic gradient descent estimators.  
	\end{IEEEbiographynophoto}
	
	\text{}\vskip -1.5 cm\text{}
	\begin{IEEEbiographynophoto}{Debasis Kundu}
		received his Ph.D. degree from
		the Pennsylvania State University, Pennsylvania, PA,
		USA, in 1989, under the guidance of Prof. C. R. Rao.
		After finishing his Ph.D., he joined The University
		of Texas at Dallas, Dallas, TX, USA, as an Assistant Professor, before joining the Indian Institute of
		Technology Kanpur, Kanpur, India, in 1990. From
		2011 to 2014, he was the Head of the Department
		of Mathematics and Statistics, IIT Kanpur, and since
		2008, an Endowed Chair Professor, and since 2019,
		the Dean of Faculty Affairs of the Institute. He has published more
		than 300 research papers in different refereed journals. He has coauthored two
		research monographs on Statistical Signal Processing and on Step-Stress Models, and coedited a book on Statistical Computing. He works on different areas of
		statistics. His major research interests include distribution theory, lifetime data
		analysis, censoring, statistical signal processing, and statistical computing. Prof.
		Kundu is the recipient of the Chandana Award from the Canadian Mathematical
		Society, Distinguished Statistician Award from the Indian Society of Probability
		and Statistics and the P. C. Mahalanobis Award from the Indian Society of
		Operation Research. He is currently the Editor-in-Chief of the Journal of the
		Indian Society of Probability and Statistics, and a member of the editorial
		boards of Communications in Statistics—Theory and Methods, Communications
		in Statistics—Simulation and Computation, and Sankhya, Series B. He was a
		member of the editorial boards of the IEEE TRANSACTIONS ON RELIABILITY,
		Journal of Statistical Theory and Practice, and Journal of Distribution Theory.
		He is a Fellow of the National Academy of Sciences, India and a Fellow of the
		Royal Statistical Society London.
	\end{IEEEbiographynophoto}%
	\text{}\vskip -2.2 cm\text{}
	\begin{IEEEbiographynophoto}{Amit Mitra}
		received his B.Sc. degree in Statistics from University of
		Calcutta. He obtained his M.Sc. degree and Ph.D. from Indian Institute
		of Technology Kanpur, both in Statistics. He is a Professor of Statistics at the Department
		of Mathematics and Statistics, Indian Institute of Technology Kanpur. His research interests
		are statistical signal processing, data mining of financial and economic
		time series and parameter estimation of non-linear time series models.
	\end{IEEEbiographynophoto}
	\begin{IEEEbiographynophoto}{Rhythm Grover}
		received her B.Sc. degree from Delhi University. She
		obtained her M.Sc. and Ph.D. from Indian Institute of Technology Kanpur. She was a postdoctoral researcher at Indian Statistical Institute in the Theoretical Statistics and Mathematics Unit .  She joined Mehta Family School of Data Science and Artificial Intelligence, IIT Guwahati as an Assistant Professor in 2021. Her research interests lie at the
		intersection of statistics and signal processing, particularly in development of efficient algorithm for parameter estimation of signal processing models, derivation of statistical properties of classical parameter estimation methods and robust estimation of model parameters in the presence of outliers.
	\end{IEEEbiographynophoto}%
\end{document}


\title{On estimating parameters of a multi-component Chirp Model with equal chirp rates}
	\author{Abhinek~Shukla, ~Debasis~Kundu,   ~Amit~Mitra, and~Rhythm~Grover
		\thanks{A. Shukla, D. Kundu  and A. Mitra are with Department of Mathematics and Statistics, Indian Institute of Technology Kanpur, Kanpur - 208016, India.}
		\thanks{R. Grover is with Mehta Family School of Data Science and Artificial Intelligence, Indian Institute of Technology Guwahati, Assam-781039, India.}
		
		\thanks{(Corresponding author: Abhinek Shukla, email: abhushukla@gmail.com)}}
	
	\maketitle

	\section{Model under study:}
	In this paper, we have considered the following real valued multi-component chirp model with equal chirp rate parameters as: \begin{align} \label{true model}
		y(n)=&\nonumber\displaystyle\sum_{k=1}^{p}\bigg(A_k^0\cos(\alpha_k^0 n+\beta^0n^2)+B_k^0\sin(\alpha_k^0 n+\beta^0n^2)\bigg)\\&+X(n),\hspace{10pt} n=1,2,3,\ldots,N ,
	\end{align} 
	where, 	\(\bm{\theta}^0=(\bm{A}^{0\top},  \bm{\xi}^{0\top})^\top\)  \(=(\underbrace{A_1^0,B_1^0,A_2^0,B_2^0,\ldots,A_p^0,B_p^0}_{\bm{A}^\top},\underbrace{\alpha_1^0,\alpha_2^0,\ldots,\alpha_p^0,\beta^0}_{\bm{\xi}^\top})^\top\) is the parameter vector, \(A_k^0, B_k^0\) are the amplitude parameters, \(\alpha_k^0\) is the frequency parameter, \(\beta^0\) is the chirp rate parameter of the \(k^{th}\) component, \(k=1,2,3,\ldots,p\) and \(X(n)\) is a noise term.
	\FloatBarrier
	\section{Figures} 
	
	\begin{figure}[h!]
		\centering
		\subfloat[\textit{Estimators of $\alpha_1^0$,  $\alpha_2^0$ and  $\alpha_3^0$. }
		]{\label{fig:alp2}{\includegraphics[width=0.6\textwidth]{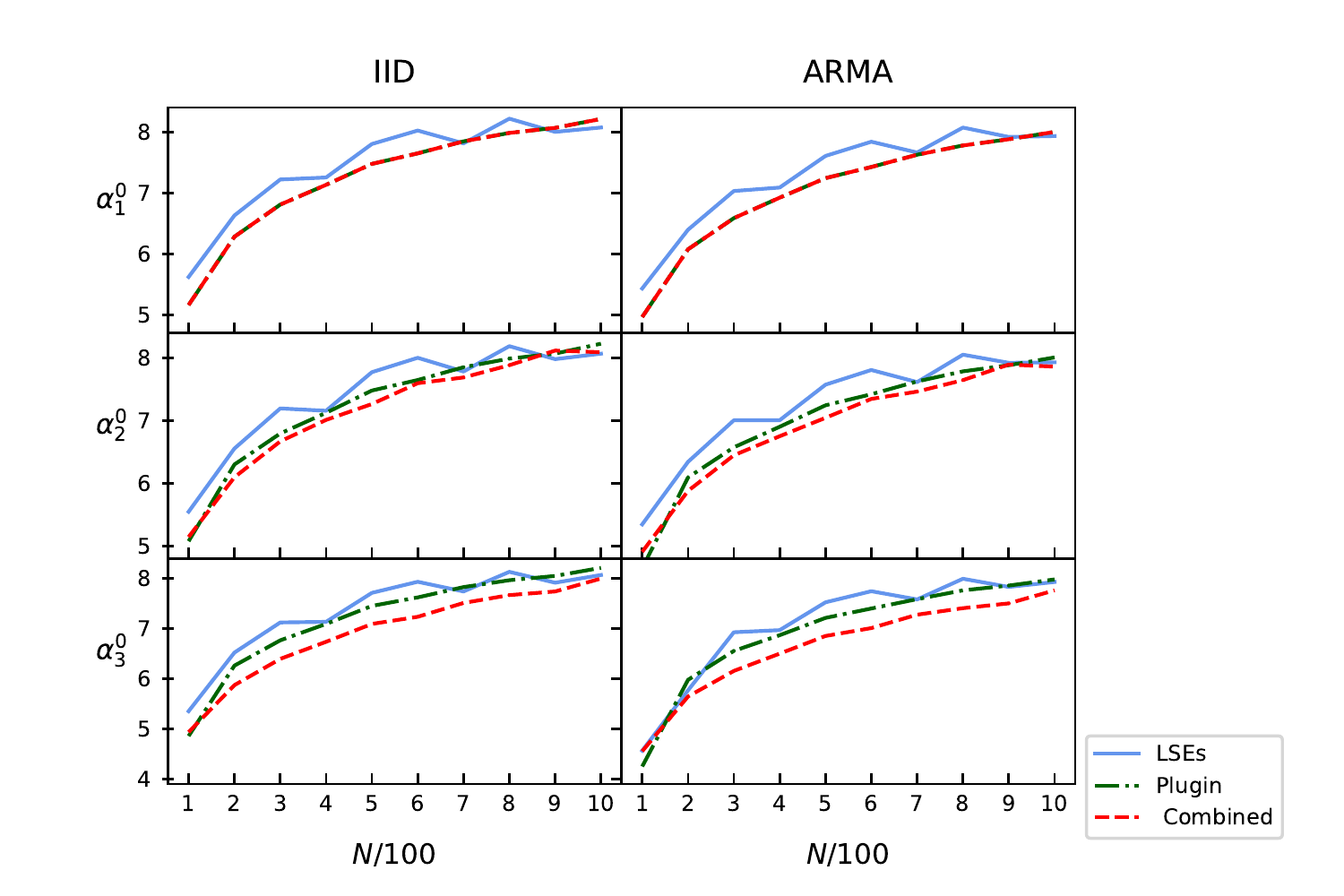}}}\qquad
		\subfloat[\textit{Estimators of $\alpha_4^0$, $\beta^0$ and $\alpha_5^0$.}]{\label{fig:beta2}{\includegraphics[width=0.6\textwidth]{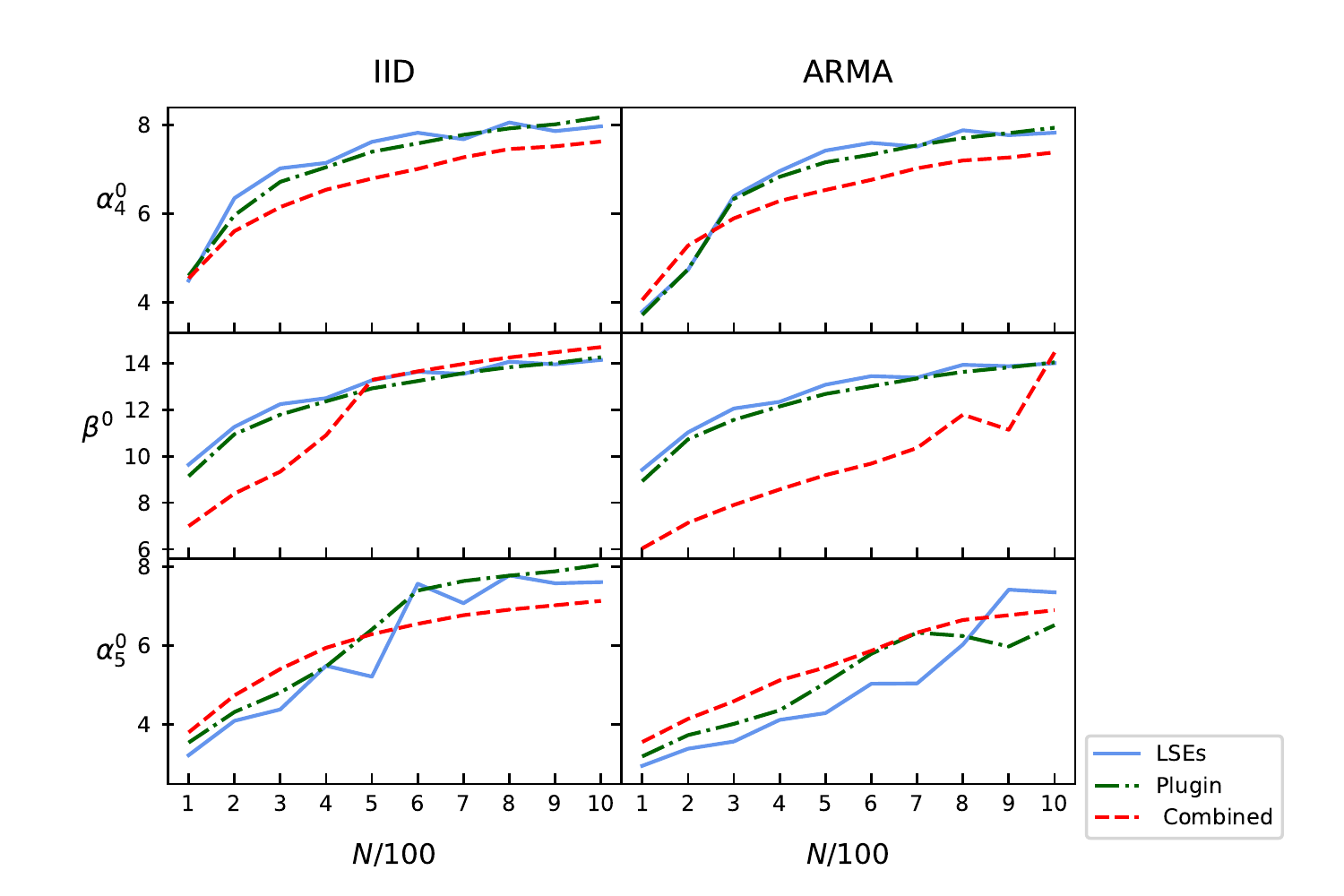}}}\\
		\caption{\textit{MSEs (on \(-\log_{10}\) scale) of LSEs, sequential plugin estimators  (Plugin) and sequential combined estimators (Combined)   versus the increasing sample size, for the case \(\sigma=3\).}}
		\label{plot_first_comp2}
	\end{figure}

\begin{figure}[htbp!]
		\centering
		\includegraphics[width=\linewidth]{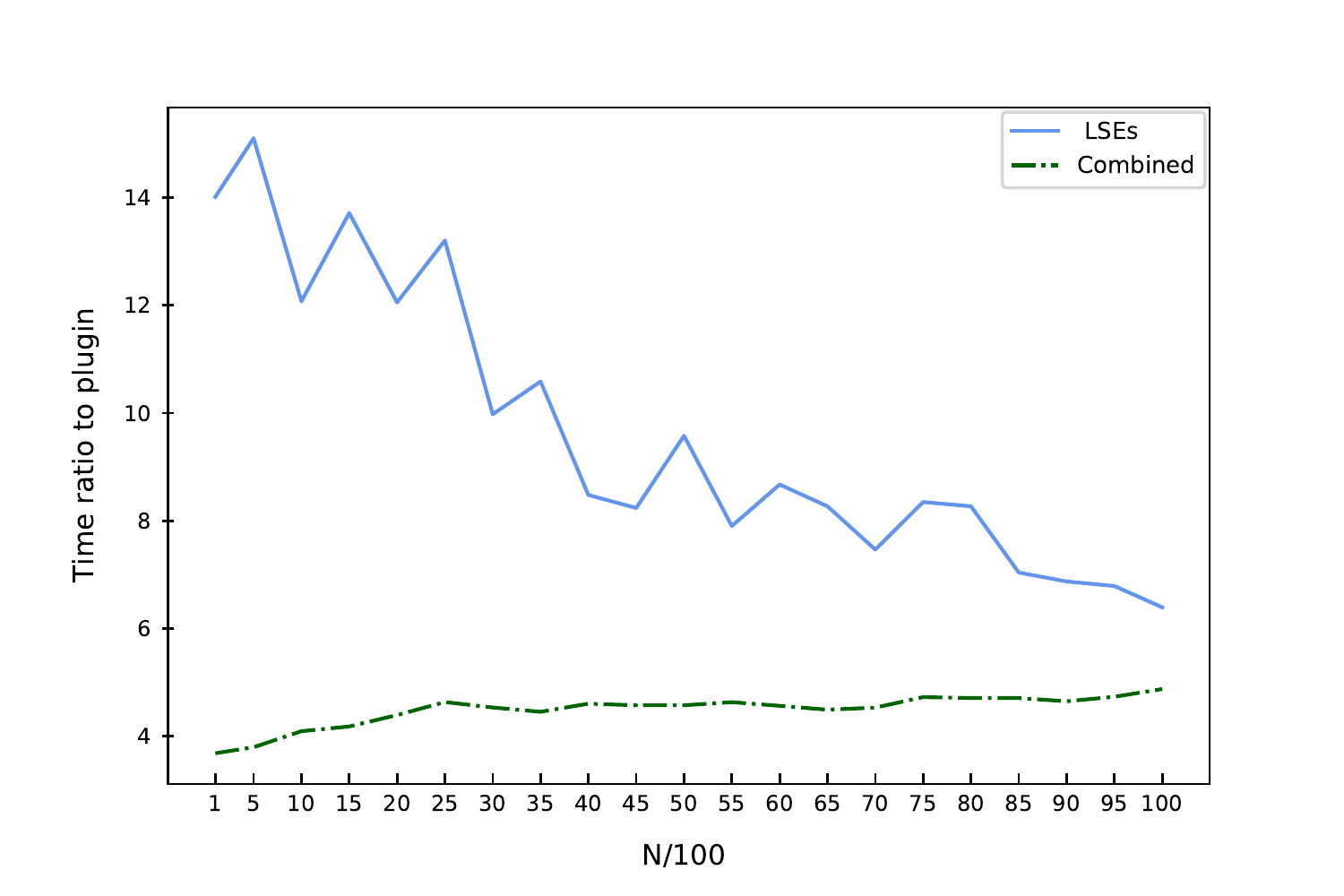}\\
		\caption{\textit{Time to compute estimators are provided in ratio to that of time required by the sequential plugin estimators.}}
		\label{plot_first_comp5}
	\end{figure}
	
	\begin{table*}										
		
		\centering									
		\begin{tabular}{|c|c|c|c|c|}									
			\hline								
			SNR	&		LSEs	&	Plug-in	&	Combined	&	ICPF	\\ \hline 
			-10	&		7.45E-13	&	2.06E-07	&	5.55E-06	&	4.03E-06	\\
			-9.5	&		6.56E-13	&	3.11E-09	&	3.82E-06	&	1.38E-06	\\
			-9	&		7.45E-13	&	2.06E-07	&	5.55E-06	&	4.03E-06	\\
			-8.5	&		4.78E-13	&	1.29E-12	&	1.88E-06	&	1.13E-07	\\
			-8	&		5.57E-13	&	1.48E-12	&	2.76E-06	&	5.62E-07	\\
			-7.5	&		3.55E-13	&	9.85E-13	&	8.34E-07	&	1.64E-12	\\
			-7	&		4.16E-13	&	1.16E-12	&	1.27E-06	&	1.69E-08	\\
			-6.5	&		2.76E-13	&	7.65E-13	&	4.01E-07	&	8.23E-13	\\
			-6	&		3.16E-13	&	8.77E-13	&	6.24E-07	&	4.89E-10	\\
			-5.5	&		2.03E-13	&	6.14E-13	&	1.76E-07	&	6.01E-13	\\
			-5	&		2.36E-13	&	6.78E-13	&	2.55E-07	&	6.96E-13	\\
			-4.5	&		1.62E-13	&	4.82E-13	&	6.72E-08	&	4.59E-13	\\
			-4	&		1.84E-13	&	5.42E-13	&	1.17E-07	&	5.38E-13	\\
			-3.5	&		1.30E-13	&	3.91E-13	&	2.65E-08	&	3.70E-13	\\
			-3	&		1.42E-13	&	4.27E-13	&	3.90E-08	&	4.10E-13	\\
			-2.5	&		1.01E-13	&	2.98E-13	&	8.34E-09	&	2.80E-13	\\
			-2	&		1.14E-13	&	3.42E-13	&	1.58E-08	&	3.14E-13	\\
			-1.5	&		8.30E-14	&	2.43E-13	&	1.53E-09	&	2.22E-13	\\
			-1	&		9.35E-14	&	2.69E-13	&	3.31E-09	&	2.53E-13	\\
			-0.5	&		6.47E-14	&	1.91E-13	&	3.33E-10	&	1.72E-13	\\
			0	&		5.74E-14	&	1.68E-13	&	1.45E-10	&	1.52E-13	\\
			0.5	&		5.12E-14	&	1.55E-13	&	1.58E-10	&	1.36E-13	\\
			1	&		4.62E-14	&	1.33E-13	&	2.02E-11	&	1.20E-13	\\
			1.5	&		4.14E-14	&	1.20E-13	&	6.14E-14	&	1.04E-13	\\
			2	&		3.71E-14	&	1.08E-13	&	2.53E-11	&	9.51E-14	\\
			2.5	&		3.32E-14	&	9.51E-14	&	4.41E-14	&	8.36E-14	\\
			3	&		2.94E-14	&	8.64E-14	&	3.81E-14	&	7.51E-14	\\
			3.5	&		2.65E-14	&	7.55E-14	&	3.40E-14	&	6.66E-14	\\
			4	&		2.41E-14	&	6.90E-14	&	3.10E-14	&	6.00E-14	\\
			4.5	&		2.06E-14	&	6.20E-14	&	2.70E-14	&	5.23E-14	\\
			5	&		1.85E-14	&	5.38E-14	&	2.41E-14	&	4.60E-14	\\
			5.5	&		1.65E-14	&	4.86E-14	&	2.11E-14	&	4.17E-14	\\
			6	&		1.53E-14	&	4.31E-14	&	1.97E-14	&	3.78E-14	\\
			6.5	&		1.36E-14	&	3.85E-14	&	1.74E-14	&	3.31E-14	\\
			7	&		1.21E-14	&	3.39E-14	&	1.58E-14	&	3.01E-14	\\
			7.5	&		1.08E-14	&	3.11E-14	&	1.39E-14	&	2.60E-14	\\
			8	&		1.00E-14	&	2.77E-14	&	1.27E-14	&	2.41E-14	\\
			8.5	&		8.75E-15	&	2.45E-14	&	1.12E-14	&	2.17E-14	\\
			9	&		7.97E-15	&	2.20E-14	&	1.03E-14	&	1.96E-14	\\
			9.5	&		7.26E-15	&	2.00E-14	&	9.29E-15	&	1.73E-14	\\
			10	&		6.44E-15	&	1.82E-14	&	8.17E-15	&	1.56E-14	\\ \hline

	\end{tabular}				
\caption{\textcolor{blue}{\textit{MSEs comparison with ICPF}}.}				
\label{mses_icpf}				
\end{table*}

	\section{Some further results}\label{consis_proof}
	
	We observe following results after some tedious calculations for $k\geq2$:
	\begin{enumerate}
		\item \begin{align*}
			AsymVar\bigg(	\cfrac{\partial Q_k(\bar{\bm{\theta}}_k^0)}{\sqrt{N}\partial{A_k}}\bigg)&=2c\sigma^2\bigg(1+\cfrac{20B_k^{0^2}}{A_1^{0^2}+B_1^{0^2}}\bigg),
		\end{align*}
		\item \begin{align*}
			AsymVar\bigg(	\cfrac{\partial Q_k(\bar{\bm{\theta}}_k^0)}{\sqrt{N}\partial{B_k}}\bigg)&=2c\sigma^2\bigg(1+\cfrac{20A_k^{0^2}}{A_1^{0^2}+B_1^{0^2}}\bigg)	,
		\end{align*}
		\item \begin{align*}
			AsymVar\bigg(	\cfrac{\partial Q_k(\bar{\bm{\theta}}_k^0)}{N\sqrt{N}\partial{\alpha_k}}\bigg)&=2c\sigma^2\bigg(\frac{A_k^{0^2}+B_k^{0^2}}{3}\\&+\cfrac{45(A_k^{0^2}+B_k^{0^2})^2}{4(A_1^{0^2}+B_1^{0^2})}\bigg)	,
		\end{align*}

		\item \begin{align*}
			AsymCoVar\bigg(	\cfrac{\partial Q_k(\bar{\bm{\theta}}_k^0)}{\sqrt{N}\partial{A_k}}&,	\cfrac{\partial Q_k(\bar{\bm{\theta}}_k^0)}{\sqrt{N}\partial{B_k}}\bigg)\\&=2c\sigma^2\bigg(-\cfrac{20B_k^{0}A_k^0}{A_1^{0^2}+B_1^{0^2}}\bigg),
		\end{align*}
		\item \begin{align*}
			AsymCoVar\bigg(&	\cfrac{\partial Q_k(\bar{\bm{\theta}}_k^0)}{\sqrt{N}\partial{A_k}},	\cfrac{\partial Q_k(\bar{\bm{\theta}}_k^0)}{N\sqrt{N}\partial{\alpha_k}}\bigg)\\&=2c\sigma^2B_k^0\bigg(1/2+\cfrac{15(A_k^{0^2}+B_k^{0^2})}{A_1^{0^2}+B_1^{0^2}}\bigg),	
		\end{align*}
		\item \begin{align*}
			AsymCoVar\bigg(&	\cfrac{\partial Q_k(\bar{\bm{\theta}}_k^0)}{\sqrt{N}\partial{B_k}},\cfrac{\partial Q_k(\bar{\bm{\theta}}_k^0)}{N\sqrt{N}\partial{\alpha_k}}\bigg)\\&=-2c\sigma^2A_k^{0}\bigg(1/2+\cfrac{15(A_k^{0^2}+B_k^{0^2})}{A_1^{0^2}+B_1^{0^2}}\bigg)	.
		\end{align*}
	\end{enumerate}
	And asymptotic covariance between $k^{th}$ and $j^{th}$ components estimators for $1<k< j\leq p,$ can be obtained by using following results:
	\begin{enumerate}
		\item \begin{align*}
			AsymCoVar\bigg(	\cfrac{\partial Q_k(\bar{\bm{\theta}}_k^0)}{\sqrt{N}\partial{A_k}},	\cfrac{\partial Q_j(\bar{\bm{\theta}}_j^0)}{\sqrt{N}\partial{A_j}}\bigg)&=2c\sigma^2\big(20 B_j^0B_k^0\big),
		\end{align*}
		
		\item \begin{align*}
			AsymCoVar\bigg(	\cfrac{\partial Q_k(\bar{\bm{\theta}}_k^0)}{\sqrt{N}\partial{A_k}},	\cfrac{\partial Q_j(\bar{\bm{\theta}}_j^0)}{\sqrt{N}\partial{B_j}}\bigg)&=2c\sigma^2\big(-20 A_j^0B_k^0\big),
		\end{align*}
		\item \begin{align*}
			AsymCoVar\bigg(	&\cfrac{\partial Q_k(\bar{\bm{\theta}}_k^0)}{\sqrt{N}\partial{A_k}},	\cfrac{\partial Q_j(\bar{\bm{\theta}}_j^0)}{N\sqrt{N}\partial{\alpha_j}}\bigg)\\&=2c\sigma^2\bigg(15B_k^0\big (A_j^{0^2}+B_j^{0^2}\big)\bigg),
		\end{align*}
		\item \begin{align*}
			AsymCoVar\bigg(	\cfrac{\partial Q_k(\bar{\bm{\theta}}_k^0)}{\sqrt{N}\partial{B_k}},	\cfrac{\partial Q_j(\bar{\bm{\theta}}_j^0)}{\sqrt{N}\partial{A_j}}\bigg)&=2c\sigma^2\big(-20 B_j^0A_k^0\big),	
		\end{align*}
		\item \begin{align*}
			AsymCoVar\bigg(	\cfrac{\partial Q_k(\bar{\bm{\theta}}_k^0)}{\sqrt{N}\partial{B_k}},	\cfrac{\partial Q_j(\bar{\bm{\theta}}_j^0)}{\sqrt{N}\partial{B_j}}\bigg)&=2c\sigma^2\big(20 A_j^0A_k^0\big),	
		\end{align*}
		
		\item \begin{align*}
			AsymCoVar\bigg(&	\cfrac{\partial Q_k(\bar{\bm{\theta}}_k^0)}{\sqrt{N}\partial{B_k}},	\cfrac{\partial Q_j(\bar{\bm{\theta}}_j^0)}{N\sqrt{N}\partial{\alpha_j}}\bigg)\\&=2c\sigma^2\bigg(-15A_k^0\big (A_j^{0^2}+B_j^{0^2}\big)\bigg)	,
		\end{align*}
		\item \begin{align*}
			AsymCoVar\bigg(&	\cfrac{\partial Q_k(\bar{\bm{\theta}}_k^0)}{N\sqrt{N}\partial{\alpha_k}},\cfrac{\partial Q_j(\bar{\bm{\theta}}_j^0)}{N\sqrt{N}\partial{\alpha_j}}\bigg)\\&=2c\sigma^2\bigg(\big(A_k^{0^2}+B_k^{0^2}\big)\big(A_j^{0^2}+B_j^{0^2}\big)\cfrac{45}{4}\bigg)	,
		\end{align*}
		\item \begin{align*}
			AsymCoVar\bigg(	&\cfrac{\partial Q_k(\bar{\bm{\theta}}_k^0)}{N\sqrt{N}\partial{\alpha_k}},\cfrac{\partial Q_j(\bar{\bm{\theta}}_j^0)}{N\sqrt{N}\partial{A_j}}\bigg)\\&=2c\sigma^2\bigg(15B_j^0\big (A_k^{0^2}+B_k^{0^2}\big)\bigg),
		\end{align*}
		\item \begin{align*}
			AsymCoVar\bigg(	&\cfrac{\partial Q_k(\bar{\bm{\theta}}_k^0)}{N\sqrt{N}\partial{\alpha_k}},\cfrac{\partial Q_j(\bar{\bm{\theta}}_j^0)}{N\sqrt{N}\partial{B_j}}\bigg)\\&=2c\sigma^2\bigg(-15A_j^0\big (A_k^{0^2}+B_k^{0^2}\big)\bigg)	.	
		\end{align*}
	\end{enumerate}
	A short comparison between asymptotic variance of sequential plugin and sequential combined estimators of   $\alpha_k^0$, $A_k^0$ and $B_k^0$, for $k\geq2$:
	
	\begin{enumerate}
		
		\item \begin{align*}
			AsymVar(\widetilde{\alpha}_k)&-AsymVar(\breve{\alpha}_k)\\&=\cfrac{360c\sigma^2\big(A_k^{0^2}+B_k^{0^2}-(A_1^{0^2}+B_1^{0^2})\big)}{(A_1^{0^2}+B_1^{0^2})(A_k^{0^2}+B_k^{0^2})}<0.
		\end{align*}
		\item Since $A_k^{0^2}+B_k^{0^2}<A_1^{0^2}+B_1^{0^2} \implies$\\$ \cfrac{1}{A_1^{0^2}+B_1^{0^2}}<\cfrac{1}{A_k^{0^2}+B_k^{0^2}}$,  and for $A_k^{0^2}\neq B_k^{0^2}$ \begin{align*}
			&\Delta_1 =	\cfrac{AsymVar(\widetilde{A}_k)-AsymVar(\breve{A}_k)}{2c\sigma^2}\\&=\cfrac{\big(3A_k^{0^2}-8B_k^{0^2}\big)}{(A_k^{0^2}+B_k^{0^2})}+\cfrac{\big(5A_k^{0^2}\big)}{(A_1^{0^2}+B_1^{0^2})}\\&\implies  \cfrac{\big(3A_k^{0^2}-8B_k^{0^2}\big)}{(A_1^{0^2}+B_1^{0^2})}+\cfrac{\big(5A_k^{0^2}\big)}{(A_1^{0^2}+B_1^{0^2})}<\Delta_1\\&< \cfrac{\big(3A_k^{0^2}-8B_k^{0^2}\big)}{(A_k^{0^2}+B_k^{0^2})}+\cfrac{\big(5A_k^{0^2}\big)}{(A_k^{0^2}+B_k^{0^2})}\\&\iff \cfrac{8\big(A_k^{0^2}-B_k^{0^2}\big)}{(A_1^{0^2}+B_1^{0^2})}<\Delta_1< 
			\cfrac{8\big(A_k^{0^2}-B_k^{0^2}\big)}{(A_k^{0^2}+B_k^{0^2})} .
		\end{align*}
		Here, \(\Delta_1< 0\) for \(A_k^{0^2}<B_k^{0^2}\).\\~\\
		\item  Similarly, for $A_k^{0^2}\neq B_k^{0^2}$, we can observe \begin{align*}
			&\Delta_2=	\cfrac{AsymVar(\widetilde{B}_k)-AsymVar(\breve{B}_k)}{2c\sigma^2}=\cfrac{\big(3B_k^{0^2}-8A_k^{0^2}\big)}{(A_k^{0^2}+B_k^{0^2})}\\&+\cfrac{\big(5B_k^{0^2}\big)}{(A_1^{0^2}+B_1^{0^2})}
		\implies \cfrac{8\big(B_k^{0^2}-A_k^{0^2}\big)}{(A_1^{0^2}+B_1^{0^2})}<\Delta_2< 
			\cfrac{8\big(B_k^{0^2}-A_k^{0^2}\big)}{(A_k^{0^2}+B_k^{0^2})} .
		\end{align*}
		
	\end{enumerate}
	Here, \(\Delta_2> 0\) for \(A_k^{0^2}<B_k^{0^2}\).